\documentstyle[preprint,tighten,epsfig,aps,mathbbol,floats]{revtex}

\def\gt{\tilde{g}}
\def\H{{\cal H}} 
\newcommand{\fd}{fluc\-tu\-a\-tion-dis\-si\-pa\-tion }
\def\be{\begin{equation}}
\def\ee{\end{equation}}

\def\bea{\begin{eqnarray}}
\def\eea{\end{eqnarray}}

\def\x{{\bf x}}
\def\q{{\bf q}}
\def\rmd{{\rm d}}
\def\rme{{\rm e}}
\def\e{\epsilon}
\def\dpar{\partial}
\def\p{\varphi}
\def\t{\tau}

\def\Li{{\rm Li}}
\def\ft{\tilde\psi}
\def\f{\psi}
\def\tm{\tau_m}
\def\ie{i.e.,}
\def\eg{e.g.,}
\def\k{\kappa}
\def\dm{\frac{d}{2}}
\newcommand{\pt}{{\tilde\varphi}}
\newcommand{\bt}{{\overline\theta}}
\newcommand{\reff}[1]{(\ref{#1})}
\def\MC{{\rm MC}}
\newcommand{\eq}{{\rm eq}}

\newcommand{\F}{{\mathcal F}}

\begin{document}

\draft
\title{Critical aging of Ising ferromagnets relaxing from an ordered state}
\author{Pasquale Calabrese${}^{1}$, Andrea Gambassi${}^{2,3}$, and Florent
  Krzakala${}^{4}$}
\address{$^1$Institute for Theoretical Physics, Amsterdam University,
  Valckenierstraat 65, 1018 XE Amsterdam, The Netherlands.}
\address{$^{2}$Max-Planck-Institut f\"ur Metallforschung, Heisenbergstr. 3,
  D-70569 Stuttgart, Germany.}
\address{$^{3}$Institut f\"ur Theoretische und Angewandte Physik,
  Universit\"at Stuttgart, Pfaffenwaldring 57, D-70569 Stuttgart, Germany.}
\address{$^{4}$Laboratoire Physico-Chimie Th\'eorique, UMR CNRS 7083, Ecole
  Sup\'erieure de Physique et Chimie Industrielles (ESPCI), 10 rue Vauquelin,
  Paris 75005, France.}

\date{\today}

\maketitle

\begin{abstract} 
We investigate the nonequilibrium behavior
of the $d$-dimensional Ising model with purely dissipative
dynamics during its critical 
relaxation from a magnetized initial configuration.
The {\it universal} scaling forms of the two-time response and
correlation functions of the magnetization are derived within
the field-theoretical approach and the associated scaling functions 
are computed up to first order in the $\e$-expansion ($\e = 4-d$). 
Aging behavior is clearly displayed and the associated {\it universal} 
\fd ratio tends to 
$X^\infty = \frac{4}{5} [1 - (\frac{73}{480} - \frac{\pi^2}{80}) 
\e + O(\e^2)]$ 
for long times. These results are 
confirmed by 
%means of 
Monte Carlo simulations of the two-dimensional Ising model with Glauber
dynamics, from which we find $X^\infty_\MC = 0.73(1)$.
The crossover to the case of relaxation from a disordered
state is discussed and the crossover function for the \fd
ratio is computed within the Gaussian approximation. 
\end{abstract}

\section{Introduction}
The dynamics of statistical systems close to critical points has been
the subject of intensive theoretical and experimental investigations in the
last three decades, during which mainly equilibrium properties have been
studied in detail. 
As in the case of static properties, the presence of a nearby critical
point greatly facilitate the study of 
dynamical behavior. In fact,
upon approaching a second-order phase transition some of the relevant
features of the dynamics (at length and time scales much larger than
the microscopic ones) are determined only by the fluctuations of a
suitable {\it order parameter} and 
they become to some extent independent of the actual microscopic
details of the system ({\it universality}).
In turn, this allows a quantitative characterization of the
dynamical behavior within the various universality classes grouping
together the {\it microscopically different} systems which display
{\it the same} critical behavior. This can be usually done to an
extent that is in general well out of reach for the
general case of
non-critical dynamics.  
% 

% Nonequilibrium behavior, slow dynamics....
Critical dynamics provides also a simple instance of slow collective
evolution.
Statistical systems with slow dynamics have recently attracted
considerable theoretical and experimental interest, in view of the
rich scenario of phenomena they display: Dramatic slowing down of
relaxation processes, hysteresis, memory effects, etc.
After a perturbation a system with slow dynamics does not generically achieve
equilibrium even for long times and its dynamics
is neither invariant under time translations nor under time reversal,
as it should be in thermal equilibrium. 
During this never-ending
relaxation {\it aging} occurs: Two-time quantities such as response
and correlation functions depend on the two times $s$ and
$t>s$ not via $t-s$ only and their decays as functions of $t$ are slower
for larger $s$. At variance with one-time quantities (e.g., the order
parameter) -- converging to asymptotic values in the long-time limit --
two-time quantities clearly bear the signature of aging.
Aging was known to occur in disordered and complex systems (see, e.g., Ref. 
\cite{review}) and only
in the last ten years attention has been focused on simpler systems
such as critical ones,
whose universal features can be rather
easily investigated by using different methods and which might provide
insight into more general cases~\cite{cg-rev} (see also Ref.~\cite{g-05}
for a pedagogical introduction).  
Indeed, consider a system with a critical point at temperature 
$T_c$, order parameter $\p({\bf x},t)$, and
prepare it in some initial configuration (which might correspond to an
equilibrium state at a given temperature $T_0$). 
At time $t=0$ bring
the system in contact with a thermal bath of temperature $T_b$($\neq T_0$). 
The ensuing relaxation process is expected to be
characterized by some {\it equilibration time} $t_{\rm eq}(T_b)$ 
such that for $t \gg t_{\rm eq}(T_b)$ equilibrium is
attained and the dynamics is stationary and invariant under
time reversal, whereas for $0 < t \ll t_{\rm eq}(T_b)$, 
the evolution depends on the specific initial condition.
Upon approaching the critical point
$T_b = T_c$ the equilibration time diverges and therefore
equilibrium is never achieved. 
To monitor the time evolution we consider the average order parameter
$M(t) = \langle \p({\bf x},t)\rangle$  (translational invariance in
space is assumed),  the time-dependent correlation
function  of the order parameter $C_{\bf x}(t,s)= \langle \p({\bf
x},t)\p({\bf 0},s)\rangle$, where $\langle \ldots \rangle$ stands
for the mean over the stochastic dynamics, and the linear
response (susceptibility) $R_{\bf x}(t,s)$ to an external field.  
$R_{\bf x}(t,s)$ is defined by
$R_{\bf x} (t,s)=\delta\langle \p({\bf x},t)\rangle/ \delta h (s)$,
where $h$ is a small external field, conjugate to $\p({\bf x}={\bf
0},s)$ (e.g., if $\p$ is a magnetic order parameter, $h$ is the
magnetic field), applied at time $s > 0$ at the point ${\bf x}={\bf
0}$. Note that causality implies $R_{\bf x}(t,s>t) = 0$ and that 
$C_{\bf x}(t,s) = C_{\bf x}(s,t)$ in the bulk. 
According to the general picture of the relaxation process, one
expects that for $t > s \gg t_\eq(T_b)$,
$C_{\bf x}(t,s) = C^\eq_{\bf x}(t-s)$
and $R_{\bf x}(t,s) = R^\eq_{\bf x}(t-s)$ where $C^\eq_{\bf x}$ and
$ R^\eq_{\bf x}$ are the corresponding equilibrium quantities, related
by the {\it \fd theorem} (FDT)
\be
R^\eq_{\bf x}(\tau > 0) = - \frac{1}{T_b} \frac{\rmd C^\eq_{\bf
x}(\tau)}{\rmd\tau} \;,
\label{eqFDT}
\ee
where the temperature is measured in unit of the Boltzmann's constant.
The FDT suggests the definition of the so-called {\it \fd ratio} (FDR)~\cite{ck-93,ckp-94}:
\be
X_{\bf x}(t,s)=\frac{T_b\, R_{\bf x}(t,s)}{\dpar_s C_{\bf x}(t,s)} \; ,
\label{dx}
\ee
where we assume $t>s$. According to the previous picture 
for $t>s\gg t_\eq(T_b)$
the FDT yields $X_{\bf x}(t,s)=1$.
This is not generically true in the aging regime.
The asymptotic value of the FDR 
\be
X^\infty=\lim_{s\to\infty}\lim_{t\to\infty}X_{{\bf x}=0}(t,s)
\label{xinfdef}
\ee
is a very useful quantity in the description
of systems with slow dynamics, since  $X^\infty=1$ whenever 
the aging evolution is interrupted and  
the system crosses over to equilibrium dynamics, 
i.e., $t_{\rm eq}(T_b)<\infty$.
Conversely 
$X^\infty\neq 1$ is a signal of an asymptotic non-equilibrium dynamics.
Moreover $X^\infty$ can be used to define an effective non-equilibrium
temperature 
$T_{\rm eff}=T/X^\infty$, which might have some features of the 
temperature of an equilibrium system, e.g., controlling the direction of heat 
flows and acting as a criterion for thermalization~\cite{ckp-97}.
Within the field-theoretical approach to critical dynamics
it is more convenient to focus on the
behavior of observables in {\it momentum} space. Accordingly
hereafter we mainly consider 
the momentum-dependent response $R_{\bf q}(t,s)$
and correlation $C_{\bf q}(t,s)$ functions, 
defined as the Fourier transform of $R_\x (t,s)$ and $C_\x(t,s)$,
respectively.
In momentum space it is natural to introduce a 
quantity that, just like $X_{\bf x}(t,s)$, ``gauges'' the distance
from equilibrium evolution~\cite{cg-02a1}:
\be
{\cal X}_{\bf q}(t,s)=\frac{T_b R_{\bf q}(t,s)}{\dpar_s C_{\bf q}(t,s)} \;.
\label{Xq} 
\ee
Note that ${\cal X}_{\bf q}(t,s)$ is {\it not} the Fourier transform of 
$X_{\bf x}(t,s)$.
The long-time limit 
\be
{\cal X}^\infty_{{\bf q}=0}\equiv 
\lim_{s\rightarrow\infty}\lim_{t\rightarrow\infty} {\cal X}_{{\bf q}=0}(t,s)
\label{eq} \;
\ee
has been used to define an effective temperature, in analogy with
$X^\infty$. Interestingly, 
it has been argued that $X^\infty= {\cal X}^\infty_{{\bf
q}=0}$~\cite{cg-02a1} (see also Ref.~\cite{sp-05}).
However, 
for quenches to the critical point ${\cal X}^\infty_{{\bf q}=0}$ 
(and therefore $X^\infty$)
depends on
the specific choice of the quantity which $R_\q$ and $C_\q$ refer
to~\cite{cg-04} and therefore it would be difficult to assign a sound
physical meaning to $T_{\rm eff}$ defined from $X^\infty$ (see 
also \cite{as-05}). 

We shall consider here the universal aspects of one of the simplest
dynamic universality class: 
Pure relaxation (Model A in the notion of Ref.~\cite{HH}) 
within the Ising (static) universality class. 
As we shall discuss later on, the lattice Ising model with Glauber dynamics
belongs to it. 
In spite of its simplicity the non-equilibrium behavior of this model
has been investigated in some detail only recently. 
In particular,
earlier works focussed primarily on general scaling properties of the
non-linear relaxation~\cite{r-76,fr-76} close to $T_c$ and on the equation of
motion describing the dynamics of the order
parameter 
in the presence of time-dependent external fields~\cite{bj-76,bej-79}. 
Two-time
quantities such as correlation and response functions 
were not discussed explicitly beyond equilibrium~\cite{bj-76}, 
although the non-equilibrium behavior of 
$R_{\q =0}(t,s)$
was implicitly encoded in the equation of motion computed 
in Refs.~\cite{bj-76,bej-79}. 
A more careful analysis of the relaxation process from a given 
initial (non-equilibrium) 
state revealed the presence of an 
additional stage of relaxation with universal features~\cite{jss-89,jan-92}
(characterized by the {\it initial-slip} exponent $\theta$,
cf.~Sec.~\ref{sec-scaling-forms}) 
which was previously overlooked. The general analysis
of Refs.~\cite{jss-89,jan-92} provided the scaling functions for
generic multi-time quantities and in particular for the order
parameter $M(t)$ and for the two-time quantities 
$R_\x(t,s)$ and $C_\x(t,s)$. On the
other hand there was no explicit analysis of the resulting aging
behavior at the critical point, which has been later carried out in 
Ref.~\cite{ckp-94,cg-02a1,gl-00c,gl-02,cg-02a2,cg-04} by computing the
two-time response and correlation functions for the case of a
relaxation from an initially disordered state with vanishing
order parameter (corresponding to a quench from high
temperature to the critical point). In particular it has been shown
that $X^\infty = 0$ for quenches below $T_c$, whereas $X^\infty$ is a
{\it universal amplitude ratio} for quenches to $T_c$, as %firstly
pointed out by Godr\`eche and Luck \cite{gl-00c}. 
Accordingly it is possible to take advantage of this universality by computing
$X^\infty$ within the field-theoretical approach to critical
dynamics. For the Ising universality class with purely relaxational
dynamics this analysis leads to $X^\infty(d=3) = 0.429(6)$, 
$X^\infty(d=2) = 0.30(5)$ in good agreement with the Monte Carlo results 
$X^\infty_\MC(d=3) \simeq 0.40$ and 
$X^\infty_\MC(d=2) = 0.340(5)$ (see Ref.~\cite{cg-rev} for details).
These investigations have been
also extended to different dynamic universality 
classes both in the
bulk~\cite{cg-rev,cg-02rim,ph-02,cg-03,sp-05,hs-04,cl-05} and at 
surfaces~\cite{p-04,cg-rev,bp-06}.

In this paper we address the complementary 
problem of the relaxation occurring 
at the critical point from an initial state with {\it non-vanishing} value
of the order parameter. Taking advantage of previous results (mainly
reported in Refs.~\cite{jan-92,bej-79}) we provide predictions for the {\it
universal} scaling forms of two-time response and correlation
functions, discussing the resulting universal aspects of aging behavior
within the field-theoretical approach to critical phenomena. 
[Within this approach some equal-time correlation functions were
previously investigated: $C_\x(t,t)$ in
Ref.~\cite{jan-92} whereas  $C_{\q=0}(t,t)$ in
Ref.~\cite{RD-96}, in the context of a detailed study of the relaxation in
finite volume -- see also Ref.~\cite{varie}.]
We also compare the provided predictions 
with the results of Monte Carlo simulations of the two-dimensional
Ising model with Glauber dynamics. 
The same problem has been recently addressed for other
universality classes~\cite{gspr-05,as-05,ft-05}.
The comparison of our findings with these recent results is presented in the 
conclusive section.

The paper is organized as follows. In Sec. \ref{sec2} we introduce the model
and its basic features. In Sec. \ref{sec-scaling-forms} we derive the general 
scaling forms for the correlation and response functions of the order parameter
after a quench from an ordered state. In Sec. \ref{sec-Gaux} we solve the model
in the Gaussian approximation. Such an approximation allows us to fully 
describe the crossover from ordered to disordered quench. Then in Sec. 
\ref{sec-onel} we present our first-order $\e$-expansion computation for
the response and the correlation functions and from them we calculate the FDR.
Sec. \ref{sec-mc} is devoted to an extensive Monte Carlo simulation of the 
two-dimensional Ising model with Glauber dynamics. 
Finally in Sec. \ref{sec-con} we summarize our results and compare them 
with the existing literature.

\section{The Model}
\label{sec2}

The simplest non-trivial model in which aging occurs 
is probably the Ising model in $d$ dimensions 
evolving with a purely dissipative dynamics 
after a quench to the critical point.  
Its Hamiltonian on an hypercubic lattice is given by
\be
\H= - \sum_{\langle {\bf xy} \rangle} {S_{\bf x}}{S_{\bf y}} \;,
\label{HON}
\ee
where ${S_{\bf x}}=\pm1$ is a spin located at the lattice site
${\bf x}$. 
The sum runs over all pairs $\langle{\bf x y}\rangle$ of nearest-neighbor 
lattice sites. 
A purely dissipative dynamics for this model proceeds by
elementary moves that amount to randomly flipping the spin $S_\x$.
The transition rates can be arbitrarily chosen provided that 
the detailed-balance condition is satisfied. 
For analytical studies the most suited are the
Glauber ones \cite{glau}, which allow exact solutions in the one-dimensional 
case \cite{lz-00,gl-00i,fins-01,mbgs-03,ms-04,hs-04}.

Despite its simplicity, the dynamics of this model is not exactly solvable 
in physical dimensions $d=2,3$. In order 
to investigate analytically the dynamical behavior 
we take advantage of the universality considering the time evolution
of a scalar field $\p({\bf x},t)$ (obtained in principle by
coarse-graining the lattice spin variables $S_\x$) with a
purely dissipative dynamics~(Model A in the notion 
of Ref.~\cite{HH}). This is
described by the stochastic Langevin equation
\be
\label{lang}
\dpar_t \p ({\bf x},t)=-\Omega 
\frac{\delta \cal{H}[\p]}{\delta \p({\bf x},t)}+\xi({\bf x},t) \; ,
\ee
where $\Omega$ is the kinetic coefficient, 
$\xi({\bf x},t)$ a zero-mean stochastic Gaussian noise with 
\be
\langle \xi({\bf x},t) \xi({\bf x}',t')\rangle= 2 \Omega \, \delta({\bf x}-{\bf x}') \delta (t-t'),
\ee
and $\cal{H}[\p]$ is the static Hamiltonian. 
Near the critical point, $\cal{H}[\p]$ may be assumed of the 
Landau-Ginzburg form
\be
{\cal H}[\p] = \int \rmd^d x \left[
\frac{1}{2} (\nabla \p )^2 + \frac{1}{2} r_0 \p^2
+\frac{1}{4!} g_0 \p^4 \right] ,\label{lgw}
\ee
where $r_0$ is a parameter that has to be tuned to a critical value 
$r_{0,c}$ in order to 
approach the critical temperature $T=T_c$ ($r_{0,c}=0$, 
within the analytical approach discussed below), and $g_0>0$ is the bare
coupling constant of the theory. 
This coarse-grained continuum dynamics is in the same 
universality class as the lattice Ising model with spin-flip 
dynamics~\cite{HH}.

Correlation and response functions of a field satisfying the
Langevin equation~\reff{lang} can be obtained by means of the 
field-theoretical action~\cite{zj,bjw-76} 
\be
S[\p,\tilde{\p}]= \int_0^\infty \rmd t \int \rmd^dx 
\left[\tilde{\p} \dpar_t\p +
\Omega \tilde{\p} \frac{\delta \mathcal{H}[\p]}{\delta \p}-
\tilde{\p} \Omega \tilde{\p}\right]\,,\label{mrsh}
\ee
where $\pt({\bf x},t)$ is an auxiliary field, conjugate to 
the external field $h$ in such a way that
${\cal H}[\p,h] = {\cal H}[\p] - \int \rmd^d x h \, \p$.
In terms of $S$,
the average over the stochastic dynamics induced by the noise 
$\xi({\bf x},t)$ of a
quantity ${\cal O}[\p]$ depending on the  order parameter field
$\p$ is given by~\cite{zj,bjw-76} 
\be
\langle {\cal O}[\p]\rangle = \int [\rmd\pt\rmd\p]\; {\cal O}[\p]\; \rme^{-
S[\pt,\p]} \;.
\ee
The effect of the external field $h$ on this average 
(denoted, for $h\neq 0$, by $\langle \ldots \rangle_h$) is accounted for by
using ${\cal H}[\p;h]$ in Eq.~\reff{mrsh}. 
As a consequence, the linear response to the field $h$ of a generic observable
${\cal O}$ is given by
\be
\left.\frac{\delta \langle {\cal O} \rangle_h}{ \delta h({\bf x},s)}
\right|_{h=0} = 
\Omega \langle \tilde\p({\bf x},s){\cal O}\rangle \,,
\label{rfield}
\ee
and hence $\pt({\bf x},t)$ is termed response field. In particular
\be
R_{\x-{\bf x'}}(t,s) = 
\left.\frac{\delta \langle \p({\bf x},t)\rangle_h}{ \delta
h({\bf x'},s)} \right|_{h=0} = 
\Omega \langle \tilde\p({\bf x'},s)\p({\bf x},t)\rangle \,.
\label{defResp}
\ee

The effect of a macroscopic initial condition 
$\p_0({\bf x})=\p({\bf x},t=0)$ may be accounted for by  
averaging over the initial configuration
with a weight $\rme^{-H_0[\p_0]}$ where \cite{jss-89}
\be
H_0[\p_0]=\int\! \rmd^d x\, \frac{\tau_0}{2}[\p_0({\bf x})-M_0]^2,
\label{idist}
\ee
which specifies an initial state with Gaussian short-range 
correlations of the order parameter fluctuations, proportional 
to $\tau_0^{-1}$ and spatially constant initial averaged order parameter 
$M_0 = \langle \p_0(\x)\rangle$ (the generalization to space-dependent
$M_0$ is straightforward).
In the experimental protocols we have in mind the system has
initially a non-vanishing value of the order parameter
(magnetization). 
This can be obtained by preparing the system
either (a) in an equilibrium low-temperature state in the absence of
external fields or (b) by applying an external field at generic
temperature. In both cases the resulting state can be described by
Eq.~\reff{lgw} with external field $h$, leading, far enough from
the critical point, to a Gaussian distribution such as Eq.~\reff{idist} 
with  $\tau_0 \sim g_0
M_0^2$ where $M_0 = (- 6 r_0/g_0)^{1/2}$ in case (a) ($r_0 <0$) and $M_0
= (6 h/g_0)^{1/3}$ in case (b).
However, within the renormalization-group (RG) approach to the problem 
it has been shown~\cite{jss-89} that $\tau_0^{-1}$ is an
irrelevant variable in the sense that $\tau_0^{-1}$
affects only the correction to the leading long-time scaling 
behavior we are interested in. 
In view of that we fix it to 
the value $\tau_0^{-1}=0$ from the very beginning of the calculation. 
Accordingly, cases (a) and (b) previously mentioned differ only in the
corrections to the leading scaling behavior.

In order to account for a non-vanishing mean value of the order
parameter $\langle \p(\x,t)\rangle = M(t)$ (we assume that $M(t)$
stays homogeneous in space after the quench)
during the time evolution, 
it is convenient~\cite{bej-79} to write the 
action~\reff{mrsh} in terms of fluctuations around $M(t)$, i.e.,
\be
\f(\x,t)=\p(\x,t)-M(t)\,,\qquad \ft(\x,t)=\pt(\x,t)\,,
\ee
so that $\langle \f(\x,t)\rangle =0$ ($\ft$ has been introduced to
make the notation uniform). The problem we shall consider is
therefore the dynamics of fluctuations of the field $\f(\x,t)$ in the
time-dependent ``background field'' provided by $M(t)$.
For later convenience we also introduce the rescaled magnetization
$m(t)$:
\be
m^2 \equiv g_0\frac{M^2}{2}\,,
\label{mresc}
\ee
so that a perturbative expansion in $g_0$ leads to a finite value for 
$m_0 \equiv m(t=0)$ in the case (a) mentioned above.
The resulting action in terms of $\f, \ft$ may be written as
\be
S=\int_{0}^\infty \rmd t \int \rmd^dx  ({\cal L}_0+{\cal L}_1+{\cal L}_2)\,,
\ee
with 
\bea
{\cal L}_0&=&-\ft \Omega \ft +\ft
\left(\dpar_t+\Omega \left[\nabla^2+r_0+m^2(t)\right]\f\right)\,,\label{L0}\\
{\cal L}_1&=&\Omega \sqrt{\frac{g_0}{2}} m(t) \ft \f^2+
\frac{\Omega g_0}{6} \ft \f^3\,,\label{L1}\\
{\cal L}_2&=&\ft\left(\dpar_t+
\Omega\left[r_0+\frac{1}{3}m^2(t)\right]\right)
\sqrt{\frac{2}{g_0}}m(t) \equiv \ft h_{\rm eff}(t)\label{L2}\,.
\eea
We split up the action $S$ so that ${\cal L}_0$ is the Gaussian part, 
${\cal L}_1$ contains the interaction vertices and ${\cal L}_2$ gives
the coupling to the effective magnetic field 
$h_{\rm eff}(t)$ acting on $\f(\q=0,t)$ and due to a nonzero
$M(t)\propto m(t)$. Note that in ${\cal L}_0$ the effect of a
non-vanishing mean value of the order parameter $m(t)$ is equivalent
to a time-dependent temperature shift: $(m(t),r_0) \mapsto
(0,r_0+m^2(t))$. Therefore, even if the system is asymptotically (for
long times) at its critical point $r_0 = r_{0,c}=0$, for finite times it is
effectively in the disordered phase. 

Following standard methods \cite{zj,bjw-76} the response and correlation 
functions may be obtained by a perturbative expansion of the functional weight
$\rme^{-(S[\p,\pt]+H_0[\p_0])}$ in terms of the
coupling constant $g_0$. 
The propagators~(Gaussian two-point functions of the 
fields $\f$ and $\ft$ in momentum space) are~\cite{jss-89}
\bea
\langle \f(\q,t) \ft(\q',t')\rangle_0&=& (2\pi)^d \delta(\q+\q') R^0_\q(t,t')\,,\\
\langle \f(\q,t) \f(\q',t')\rangle_0&=& (2\pi)^d \delta(\q+\q')  C^0_\q(t,t')\,,
\eea
with ($q=|{\bf q}| $)
\bea
R^0_{\bf q}(t,t')&=& \theta(t-t') \exp\left\{ -\Omega\left[(q^2+r_0)(t-t')+
\int_{t'}^t \rmd t'' m^2(t'')\right]\right\}\,,\label{Rgaux}\\
C^0_{\bf q}(t,t')&=&2\Omega \int_0^\infty \rmd t'' R^0_{\bf q}(t,t'')
R^0_{\bf q}(t',t'').\label{Cgaux}
\eea
In the following we will assume the \^Ito prescription 
(see, \eg\ Refs.~\cite{jan-92,zj}) to deal with the 
ambiguities arising in formal manipulations of stochastic equations. 
Consequently, all the diagrams with loops of response propagators have 
to be omitted. 
This ensures that causality holds in the perturbative 
expansion~\cite{jss-89,jan-92,bjw-76}.
From the technical point of view, the breaking of time-translation invariance
does not allow the factorization of connected correlation functions in terms 
of one-particle irreducible ones as usually happens when
time-translation invariance holds.
As a consequence, as in the case of surface
critical phenomena \cite{diehl-86},  the whole calculation
has to be done in terms of connected functions only \cite{jss-89}.

The perturbative expansion can be as usual 
organized in terms of Feynman diagrams with 
propagators given by Eqs.~\reff{Rgaux} (represented as a directed
line with the arrow pointing to $t$, the larger of the two times $t,t'$)
and \reff{Cgaux} (represented as an unoriented line) and
vertices given by ${\cal L}_1$. In addition to the
standard time-independent quartic vertex $-\Omega g_0$ a
time-dependent cubic one $-\Omega\sqrt{2g_0} m(t)$, due to a non-zero
magnetization $m(t)$, has to be accounted for in the expansion. As
explained below, the contribution of $h_{\rm eff}(t)$ cancels for a
suitable choice of $m(t)$.
The evolution equation for the magnetization $m(t)$ can be obtained by
solving the equation of motion  $\langle \delta
S/\delta \ft(\x,t)\rangle = 0$. 
Taking into account that causality
implies $\langle \ft(\x,t)\rangle = 0$ (corresponding to 
$
\begin{minipage}{0.4in}
\centering\includegraphics[width=0.4in]{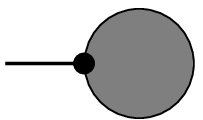}
\end{minipage} = 0
$) and that $m(t)$ has been
defined so that $\langle \f(\x,t)\rangle = 0$, one finds (in zero
external field)~\cite{bej-79}
\be
\sqrt{\frac{2}{g_0}}\left(\dpar_t+ \Omega\left[r_0+\frac{1}{3}m^2(t)\right]\right) m(t)  +
\Omega\sqrt{\frac{g_0}{2}} m(t) \langle \f^2\rangle + \frac{\Omega
g_0}{6}\langle \f^3\rangle = 0\;,
\label{eqmg}
\ee
which has to be supplemented by the proper initial condition, e.g., by
assigning the value of $m_0=m(t=0) \propto M_0$.
Note that the sum of the last two terms in this equation equals
$-{\cal T}_1$, where 
${\cal T}_1 =
\begin{minipage}{0.4in}
\centering\includegraphics[width=0.4in]{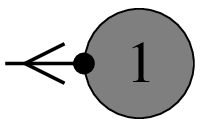}
\end{minipage} 
$ is the one-point vertex due to the
interaction vertices in ${\cal L}_1$, whereas the first term is
$-{\cal T}_2$, where 
${\cal T}_2 = 
\begin{minipage}{0.35in}
\centering\includegraphics[width=0.35in]{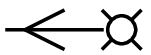}
\end{minipage}
$
is the contribution of the effective field $h_{\rm eff}(t)$ [see
Eq.~\reff{L2}] coming from ${\cal L}_2$. 
Accordingly, Eq.~\reff{eqmg} states that ${\cal T}_1
+ {\cal T}_2 = 0$, \ie\ 
$
\begin{minipage}{0.4in}
\centering\includegraphics[width=0.4in]{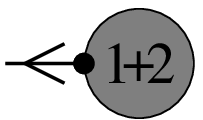}
\end{minipage}
=\!\!
\begin{minipage}{0.6in}
\centering\includegraphics[width=0.4in]{TP.eps}
\end{minipage}\!\!
+
\begin{minipage}{0.35in}
\centering\includegraphics[width=0.35in]{TPh.eps}
\end{minipage}
= 0
$
where 
$
\begin{minipage}{0.4in}
\centering\includegraphics[width=0.4in]{TP12.eps}
\end{minipage}
$ is the total one-point vertex  with one external
$\ft$-leg due to ${\cal
L}_1+{\cal L}_2$. In view of this simplification we will
omit, in the following computation, all the diagrams having the
one-point vertex as a subdiagram.
To the lowest order [hereafter $(\rmd q)=\rmd^d q/(2\pi)^d$ and $\Omega=1$]
\be
\langle \f^2(\x,t) \rangle = \int (\rmd q) C^0_\q(t,t) + O(g_0)\;,
\quad\quad 
\langle \f^3(\x,t)\rangle = O(\sqrt{g_0})
\ee
and therefore up to one loop and at the critical point $r_0 = 0$, 
Eq.~\reff{eqmg} becomes
\be
0 = \dpar_t m(t) +\frac{\Omega}{3} m^3(t) + \frac{\Omega g_0}{2}\int
(\rmd q) C^0_\q(t,t) + O(g_0^2) \,.
\label{eqmo}
\ee

\section{Scaling forms}
\label{sec-scaling-forms}

The non-equilibrium evolution of a critical system in the absence of an 
order-parameter background is a well-understood topic (see, 
e.g., Ref.~\cite{cg-rev}). In particular, it is well-known that zero-momentum 
response and correlation functions satisfy the scaling 
forms \cite{jss-89,cg-rev}
\bea
R_{{\bf q}=0}(t,s) &=& A_R\, (t-s)^a(t/s)^\theta {\cal F}_R(s/t)\; ,
\label{scalRm0}\\
C_{{\bf q}=0}(t,s) &=& A_C\,s(t-s)^a(t/s)^\theta {\cal F}_C (s/t)\; ,
\label{scalCm0}
\eea
where $a = (2-\eta-z)/z$, $z$ is the dynamical critical exponent, $\eta$  
the anomalous dimension of the field, and $\theta$ is a genuine
non-equilibrium exponent \cite{jss-89}. $A_R$ and $A_C$ are
non-universal amplitudes which are fixed by the condition ${\cal
F}_{R,C}(0)=1$. With this normalization ${\cal F}_{R,C}$ are universal.
From these scaling forms the
universality of $X^\infty$ follows as an amplitude ratio.
Although in what follows we focus mainly on the case $\q=0$, the
generalization of the  presented scaling forms to non-vanishing $\q$
amounts to the introduction of an additional scaling variable 
$y = A_\Omega \Omega q^z (t-s)$. $A_\Omega$ is a dimensional
non-universal constant which can be fixed according to some specified 
condition. 

These behaviors are clearly changed if a 
critical system evolves starting from a state with 
$M_0\neq0$.
It is known from general scaling and RG arguments that the magnetization
satisfies the scaling form \cite{jss-89}
\be
m(t) = A_m m_0 t^{a+\theta} \F_M(B_m m_0 t^\k)
\label{scalm0}
\ee
where $\k = \theta + a + \beta/(\nu z)= \theta + \beta\delta/(\nu z)$
and standard notation for critical exponents has been used 
(note that 
$\kappa$ is generically expected to be positive).
The universal scaling function 
$\F_M(v)$ has an analytic expansion around $v=0$, and, due to the
obvious symmetry $(m_0,m(t)) \mapsto (-m_0,-m(t))$, $\F_M(-v)=\F_M(v)$. 
Accordingly the 
non-universal amplitudes $A_m$ and $B_m$ can be determined, \eg\ 
by imposing $\F_M(0)= 1$ and $\F''_M(0)= -1$.
From Eq.~\reff{scalm0} one sees that the effect of a non-vanishing
initial magnetization is the introduction of an additional macroscopic 
time scale $\tau_m$ into the problem, \ie\ $\tau_m = (B_m
m_0)^{-1/\k}$ and of an additional associated scaling variable
$u\equiv t/\tau_m$ in the scaling forms.
In the long-time limit $u\gg 1$ 
one has to recover the well-known behavior of the 
critical relaxation $m(t)\sim t^{-\beta/\nu z}$ \cite{fr-76}, 
and therefore $\F_M(v \gg 1) \sim v^{-1}$. In that limit
\be
m(t) = \frac{A_m}{B_m} t^{-\beta/\nu z} \F_{M,\infty}((B_m m_0)^{1/\k} t)
\label{scalminf}
\ee
where the universal scaling function 
$\F_{M,\infty}(u) = {\cal M}_0^{(\infty)} + {\cal
M}_{-1}^{(\infty)} u^{-1} + O(u^{-2})$. 

Having understood the presence of a new scaling variable $u = (B_m
m_0)^{1/\k} t = t/\tau_m$ (alternatively $v=B_m m_0 t^\k = u^\k$), it is 
trivial to generalize the scaling forms for the response Eq.~\reff{scalRm0}
and {\it connected} correlation functions Eq.~(\ref{scalCm0}) to the case of a 
non-vanishing $m_0$:
\bea
R_{{\bf q}=0}(t,s) &=& A_R\, (t-s)^a(t/s)^\theta F_R(s/t,B_m m_0 t^\k)\; ,
\label{scalR}\\
C_{{\bf q}=0}(t,s) &=& A_C\,s(t-s)^a(t/s)^\theta F_C (s/t, B_m m_0 t^\k)\;,
\label{scalC}
\eea
where no new non-universal amplitudes have been introduced and obviously
$F_R(x,0)= {\cal F}_R(x)$ and $F_C(x,0)={\cal F}_C(x)$. 
The resulting functions $F_R$ and $F_C$ are then universal.

According to 
Eqs. (\ref{scalR}) and (\ref{scalC}) a non-vanishing mean value of the 
initial magnetization $m_0\neq 0$ affects the scaling properties of
the response and correlation function 
as soon as $B_m m_0 t^\k \sim 1$ (\ie\ $t\sim \tau_m$) 
and in particular this happens
in the long-time limit we are interested in, characterized by
$t\gg s\gg\tau_m$.
This formally corresponds to the case $m_0\rightarrow \infty$, as
opposed to the case previously considered, $m_0=0$.
In this limit one expects the scaling forms~\reff{scalR} and~\reff{scalC} to
turn into:
\bea
R_{{\bf q}=0}(t,s) &=& \bar A_R\, (t-s)^a(t/s)^\bt \bar \F_R(s/t)\; ,
\label{scalR2i}\\
C_{{\bf q}=0}(t,s) &=& \bar A_C\,s(t-s)^a(t/s)^{\bt'} \bar \F_C (s/t)\; ,
\label{scalC2i}
\eea
which resemble those for $m_0=0$. 
As before $\bar A_{R,C}$ are non-universal constants 
which are fixed by requiring $\bar \F_{R,C}(0) =1$, where $\bar \F_{R,C}(x)$
are universal functions related to the large-$v$ behavior of
$F_{R,C}(x,v)$. 
The ``new'' exponents $\bt$ and $\bt'$ are clearly
different from $\theta$.
In fact, for the related problem of the response and correlations of
fluctuating modes transverse to the direction of the decaying
magnetization in $O(N)$ models,
it has been argued~\cite{ft-05} that $\bt = \bt' =
-\beta/(\nu z)$. On the other hand, nothing is known, in general, about the 
longitudinal fluctuations that are the only degrees of freedom of the
Ising model. In particular it is not obvious, a priori, whether  $\bt$ and
$\bt'$ should be expected to be the same, if they are novel  exponents
-- as $\theta$ -- or just combinations of known ones. The rest of this
section is devoted to showing that Eqs.~\reff{scalR2i}
and~\reff{scalC2i} hold for Model A with
\be
\bt = \bt' = - \frac{\beta\delta}{\nu z} .
\label{concl}
\ee 

In order to prove Eq.~\reff{concl} 
we generalize the RG treatment of
Refs.~\cite{jss-89,jan-92}, confirming also Eqs.~(\ref{scalR})
and~(\ref{scalC}).
The general scaling properties of the connected correlation function 
$G_{n,\tilde{n},n_0}\equiv\langle[\p]^n[\pt]^{\tilde{n}}[\pt_0]^{n_0}\rangle_c$
[where $\pt_0(\x)\equiv \pt(\x,t=0)$] at the RG fixed-point and in the
presence of a non-vanishing initial magnetization (we restrict attention to 
the critical point, with $\tau_0^{-1}=0$) are given by \cite{jan-92}
\be
G_{n,\tilde{n},\tilde{n}_0}(\{\x,t\}; m_0) = 
l^{\delta(n,\tilde{n},\tilde{n}_0)}
G_{n,\tilde{n},\tilde{n}_0}(\{l\x,l^zt\}; m_0 l^{-z\k}) \label{genscal}\,,
\ee
where $\delta(n,\tilde{n},\tilde{n}_0) = n (d-2+\eta)/2 + \tilde{n}
(d+2+\tilde{\eta})/2 +  \tilde{n}_0 (d+2+\tilde{\eta} +\eta_0)/2$,
$(\tilde{\eta} -\eta)/2 = z- 2$, and $\tilde{\eta}_0 = - 2 z \theta$.
In order to 
determine the behavior of the correlation and response functions for
$s/t\ll 1$ we perform a short-distance expansion in the time variable.
Indeed, for $s\rightarrow 0$, one expects
\be
\pt(\x,s\rightarrow 0) \sim Q(s,m_0) \pt_0(\x)\,,
\ee
and therefore, inserting this relation into a generic correlation
function $G_{n,\tilde{n},\tilde{n}_0}$, one gets from Eq.~\reff{genscal}
\be
Q(s,m_0) = s^{-\theta}{\cal Q}(s m_0^{1/\k})\,,
\label{SDEQR}
\ee 
where ${\cal Q}(x)$ is finite for $x=0$~\cite{jss-89}, whereas the
behavior for $x\rightarrow\infty$ has to be determined.
According to the previous equation,
\be
V R_\q(t,s\to 0)=\langle\p(\q,t)\pt(-\q,s\to 0)\rangle \sim s^{-\theta}
{\cal Q}(s m_0^{1/\k}) 
\langle\p(\q,t)\pt_0(-\q)\rangle\,,
\label{Rint}
\ee
where the volume $V$ of the system has been introduced as a
regularization of $(2\pi)^d\delta(\q=0)$ following from translational
invariance.
Now we recall that  $\pt_0(\q=0)$ is the field conjugate to
$M_0\propto m_0$
(see Ref.~\cite{jss-89}), thus 
\be 
\langle\p(\q=0,t)\pt_0(\q=0)\rangle = \frac{\delta\langle \p(\q=0,t)\rangle}{\delta M_0}\,,
\label{dmdm0} 
\ee
where, by definition, $\langle\p(\q=0,t)\rangle = V M(t)
\propto V m(t)$, which 
scales according to Eqs.~\reff{scalm0} and~\reff{scalminf}.
Accordingly, 
the leading behavior in
Eq.~\reff{dmdm0} is  
\be
\frac{\delta m(t)}{\delta m_0} \sim \left\{
\begin{array}{lcl}
 t^{a +\theta} & \mbox{for} & t \ll \tau_m = (B_m m_0)^{-1/\k} \;,\\
 t^{-1-\beta/(\nu z)} m_0^{-1 - 1/\k}&
\mbox{for} &  t \gg \tau_m \;,
\end{array}
\right.
\ee
Consider, first, the case $s, t \ll \tau_m$ (corresponding to
$m_0\rightarrow 0$, \ie\ $\tau_m \rightarrow\infty$). 
From the previous equations we recover the known result
\be
R_{\q=0}(t,s\rightarrow 0) \sim t^a\left(\frac{t}{s}\right)^\theta \quad
(m_0\rightarrow 0)\;.
\ee
In the opposite limit  $\tau_m \ll s, t$ (corresponding to
$m_0\rightarrow \infty$, \ie\ $\tau_m\rightarrow 0$) 
one gets $R_{\q=0}(t,s\rightarrow 0) \sim
s^{-\theta}{\cal Q}(s m_0^{1/\k}) t^{-1-\beta/(\nu z)} m_0^{-1 -1/\k}$. 
Given that $m_0^{-1}$ is irrelevant by dimensional analysis, we have that for 
$m_0\to\infty$ the response function must not depend on $m_0$.
Therefore ${\cal Q}(x\rightarrow \infty)\sim x^{\k +1}$ and
\be
R_{\q=0}(t,s\rightarrow 0) \sim
t^a\left(\frac{s}{t}\right)^{1+\beta/(\nu z)+a} \quad
(m_0\rightarrow \infty,\  \mbox{\ie}\ m_0 \gg B_m^{-1} s^{-1/\k}) \;.
\ee
According to this result we find for $m_0\rightarrow \infty$
the scaling form Eq. (\ref{scalR2i}) where
\be
\bt=-\left(1 + a + \frac{\beta}{\nu z}\right)=
-\frac{\beta\delta}{\nu z}\,,
\label{thetabar}
\ee
as anticipated. Note that by  definition, we have
\be
F_R(x,v\rightarrow \infty) = \frac{\bar A_R}{A_R} x^{\theta-\bt} \bar
\F_R(x)\,, 
\label{FRinf}
\ee
and so the ratio ${\cal R}_R \equiv \bar A_R/A_R$ is universal, because it is 
the first term of the expansion of a universal function.

Let us consider now the case of the correlation function, \ie\
the short-distance expansion for the field $\p(\x,s\rightarrow 0)$. 
For $m_0 = 0$ it is of the form 
$\p(\x,s\rightarrow 0) \sim Q_1(s)\partial_s \p(\x,s)|_{s=0}$, 
given that the Dirichlet boundary conditions implies
$\p(\x,0) = 0$ when inserted into correlation functions. 
Then one can prove diagrammatically that
$\dot{\p}_0(\x)\equiv\partial_s \p(\x,s)|_{s=0} = 2 \pt_0(\x)$
when inserted into correlation functions. By means of this identity
and of Eq.~\reff{genscal} one determines the
scaling properties of $ Q_1(s)$, leading to the expected scaling
for the two-time correlation function (see Eq.~\reff{scalCm0}). 
In the present case, however, one has to take into account the
non-vanishing initial value of the magnetization. Accordingly we
expect an expansion of the form
\be
\p(\x,s\rightarrow 0) \sim Q_0(s,m_0) {\mathbb 1} +
Q_1(s,m_0)\dot{\p}_0(\x) \,,
\label{Qi}
\ee
where ${\mathbb 1}$ is the identity operator and
the scaling form of the coefficients $Q_i$ can be determined by
inserting Eq.~(\ref{Qi}) into Eq.~(\ref{genscal}). In particular this
leads to $Q_0(s,m_0) = s^{-\beta/(\nu z)}{\cal Q}_0(s m_0^{1/\k})$. As a
consequence we have for the {\it connected}
correlation function (the space dependence is understood)
\bea
C(t,s\rightarrow 0) &=& \langle \f(t)\f(s\rightarrow 0) \rangle = \langle\p(t)\p(s\rightarrow 0)\rangle - \langle
\p(t) \rangle \langle\p(s\rightarrow 0) \rangle  \nonumber \\
&\sim& Q_1(s,m_0) \left[ \langle\p(t)\dot{\p}_0\rangle - \langle
\p(t) \rangle \langle\dot{\p}_0 \rangle \right]\,.
\label{SDEm}
\eea
Note that also in the case of non-vanishing initial magnetization the
connected correlation function $C^0_\q(t,t')$ [see Eq.~\reff{Cgaux}] 
satisfies the Dirichlet
boundary condition [\ie\ $C^0_q(t,t'=0) =
C^0_\q(t=0,t')=0$]. Together with causality this implies, as in the
case $m_0 = 0$, that $\dot{\p}_0(\x)= 2 \pt_0(\x)$ when inserted
into {\it connected} correlation functions, as can be verified by going 
through the
diagrammatic proof presented in  Ref.~\cite{jss-89}. Therefore, by
inserting~\reff{SDEm} into correlation functions one finds
\be
 Q_1(s,m_0) = s^{1-\theta} {\cal Q}_1 (s m_0^{1/k})\,,
\label{SDEQ1}
\ee
where ${\cal Q}_1(x)$ is finite for $x=0$. As before, the behavior for
$x\rightarrow \infty$ has to be determined. Taking into account that
\be
V C_{\q=0}(t,s\to 0)=\langle\p(\q=0,t)\p(\q=0,s\to 0)\rangle \sim s^{1-\theta}
{\cal Q}_1(s m_0^{1/\k}) 
\langle\p(\q=0,t)\pt_0(\q=0)\rangle\,,
\label{Cint}
\ee
the reasoning proceeds as before (the only difference with
Eq.~\reff{Rint} is the exponent of the prefactor $s^{1-\theta}$),
leading to the conclusion that ${\cal Q}_1(x\rightarrow \infty)\sim
x^{\k+1}$. Accordingly in the limit $m_0 \rightarrow \infty$ one finds
\be
C_{{\bf q}=0}(t,s) = \bar A_C\, s (t-s)^a(t/s)^{\bt} \bar \F_C(s/t)\; ,
\label{scalC2}
\ee
where
\be
F_C(x,v\rightarrow \infty) = \frac{\bar A_C}{A_C} x^{\theta-\bt} \bar\F_C(x)\,,
\label{FCinf}
\ee
with universal ratio ${\cal R}_C \equiv \bar A_C/A_C$.
Interestingly enough the exponent $\bt$ in Eq.~\reff{scalC2}
is the same as the one in Eq.~\reff{scalR2i}, a result that could not
have been inferred solely on the basis of scaling arguments applied to
Eqs.~\reff{scalR} and~\reff{scalC}. 

Comparing Eqs.~\reff{scalR} and~\reff{scalC} for $m_0=0$ and 
Eqs.~\reff{scalR2i} and~\reff{scalC2i} (referring to the case
$m_0\rightarrow \infty$) one sees a 
fundamental difference between the quench from a disordered state
and from the ordered one: In the latter case {\it no}
novel exponent characterizes the aging properties and 
the non-equilibrium behavior is completely described in terms of
equilibrium exponents.

As a consequence of the scaling forms
Eqs.~\reff{scalR2i} and~\reff{scalC2i} the limiting FDR can be written as
\be
X^\infty(m_0\neq0)=\frac{{\bar A_R}}{{\bar A_C}(1-{\bar \theta})}=
\frac{{\cal R}_R (1-\theta)}{{\cal R}_C (1-{\bar \theta})} X^\infty(m_0=0)\,,
\ee
which turns out to be a {\it universal} amplitude ratio related to
$X^\infty(m_0=0)$. In Ref.~\cite{gspr-05} the question of the
universality of the FDR for ferromagnetic models has been addressed
and in particular $X^\infty(m_0\neq 0)$ has been computed for some
effectively Gaussian models (see also Sec.~\ref{sec-Gaux}). In
particular, the fact that $X^\infty(m_0\neq 0)\neq X^\infty(m_0= 0)$
has been interpreted as a signal of the existence of {\it different}
dynamic universality classes for the ``critical coarsening process'' 
depending on $m_0$ being zero or not. Although it can be considered a
semantic distinction, our point of view here is slightly different:
$X^\infty(m_0\neq 0)$ is as universal as $X^\infty(m_0=0)$ in the
sense that both do not depend on the microscopic details of the
systems
(\ie\ on the specific realization of the universality class)
but only on general properties such as symmetries, dimensionality, and
conservation laws. Instead, the difference between these two quantities is
due to the fact that the dynamic properties of a system belonging to a given
universality class depend on the additional scaling variable $u=t/\tau_m$ 
in a universal way. In turn, for long
times, $u$ can be either zero ($\tau_m\sim m_0^{-1/\kappa} 
\rightarrow\infty$, \ie\ $m_0 = 0$) or
infinite ($\tau_m$ finite, \ie\ $m_0\neq 0$), making the specific
value of $m_0^{-1}$ irrelevant (as expected from RG arguments).
We will confirm this important feature explicitly within the Gaussian 
approximation in the following section.

Note that the FDR could have been defined 
by using the {\it non-connected} 
correlation function $C^{\rm nc}_{\q}(t,s)$ instead of the connected
one $C_\q(t,s)$,
since in equilibrium both of them satisfy FDT. 
In this case we have
\be
C^{\rm nc}_\q(t,s) = C_\q(t,s) + V M(t)M(s)\,,
\ee
where $V$ is the volume of the system, leading to ($t>s$)
\be
C^{\rm nc}_{\q=0}(t,s\gg\tau_m) 
= \bar A_C\, s (t-s)^a(t/s)^{\bt} \bar \F_C(s/t)+
V C_M^2 t^{-\beta/(\nu z)}s^{-\beta/(\nu z)}\,,
\ee
with $C_M= (2/g_0)^{1/2} 
A_m{\cal M}^{(\infty)}_0/B_m$ [see Eq. (\ref{scalminf})].
In a finite but large enough volume $V$ (i.e., $V\gg \xi(t)^d \sim t^{d/z}$) 
and in the aging limit $t\gg s\gg\tau_m = (B_m m_0)^{-1/\k}$, 
the second term dominates since 
$\beta/(\nu z)<\beta\delta/(\nu z) -a$. As a consequence the FDR 
takes the trivial value $X^\infty_{\rm nc}(m_0\neq0)=0$.

\section{Gaussian Approximation}
\label{sec-Gaux}

A first (oversimplified in physical dimensions $d=2,3$) description of the 
dynamics of the model just introduced is given by the Gaussian
approximation of the action $S$ [see
Eq.~\reff{mrsh}] in which one keeps only the linear and
quadratic terms in the fields $(\f,\ft)$, neglecting additional 
anharmonic terms:
$S_G[\f,\ft] = \int_0^\infty\!\!\rmd t\int\rmd^dx({\mathcal L}_0 +
{\mathcal L}_2)$.
This amounts to setting $g_0=0$, but only after having 
performed the rescaling given by Eq.~\reff{mresc}.
In spite of its simplicity, the Gaussian approximation gives exact results 
above the upper critical dimension, equal to $4$ in Ising systems.
Consequently the result of this section have to be equivalent to those 
obtained for the fully-connected lattice ($d=\infty$) in Ref.~\cite{gspr-05}.

For $g_0=0$ and at the critical point $r_0=0$,
the equation of motion (\ref{eqmo}) simplifies to
\be
\dpar_t m_G(t) =-\frac{1}{3} m_G^3(t)\,,
\ee
whose solution is 
\be
\label{mgauss}
m_G^2=\frac{1}{\frac{2}{3} t +\frac{1}{m^2_0}}  = 
\left\{ 
\begin{array}{l}
(2 t/3)^{-1} \left(1 +
\frac{3}{2 m_0^2 t}\right)^{-1}\,, \\
m_0^2  \left(1 +
\frac{2 m_0^2 t}{3}\right)^{-1}\,.
\end{array}
\right.
\ee
This expression agrees with the scaling behavior~\reff{scalm0} 
and~\reff{scalminf}, given that,
within the Gaussian model, $\theta= a = 0$ whereas $\k = \frac{1}{2}$. 
The non-universal amplitudes appearing in Eq.~(\ref{scalm0}) are 
$A_m=1$ and $B_m=\sqrt{2/3}$ and therefore $\tau_m = (B_m m_0)^{-1/k}
= 3/(2m_0^2)$, leading to $m_G(t) = m_0(1+t/\tm)^{-1/2}$.
For $t \gg \tau_m$, $m_G(t) \sim (2 t/3)^{-1/2}$ and therefore a nonzero 
$m_0^{-1}$ affects only the corrections to the leading scaling behavior. In
this sense $m_0^{-1}$ is irrelevant for large times.

Using Eqs.~\reff{Rgaux} and~\reff{mgauss} the response function is ($t>s$)
\be
R^0_\q(t,s) = \rme^{-\q^2(t-s)-\int_s^t \rmd t' m_G^2(t')}=
\left(\frac{s+\tm}{t+\tm}\right)^{3/2}\rme^{-\q^2(t-s)}\,,
%\longrightarrow_{\hspace{-0.6cm} q\to 0}
%\left(\frac{s}{t}\right)^{3/2} \,,
\label{Rgauxe} 
\ee
and the {\it connected} correlation function
\be
C^0_\q(t,s)=2 \int_0^s \rmd t' R^0_\q(t,t')R^0_\q(s,t')=
2 \frac{\rme^{-\q^2(t+s)}}{[(t+\tm)(s+\tm)]^{3/2}} \int_0^{s} \rmd t'
(t'+\tm)^3 \rme^{2\q^2 t'}\,.
%\longrightarrow_{\hspace{-0.6cm} q\to 0}
%\frac{1}{2} s \left(\frac{s}{t}\right)^{3/2} \,,
\label{Cgauxe}
\ee
%where we have introduced $\tm = 3/(2m_0^2)$ 
%[so that $m(t) = m_0(1+t/\tm)^{-1/2}$]. 
Note that $R_\q^0(t,s) \le R^0_\q(t,s)|_{\tau_m = \infty}$, \ie\ the
system evolving from a configuration with $m_0\neq 0$ is stiffer than
in the case $m_0=0$ and also correlations are reduced, in agreement
with the fact that the system is slightly displaced from its
critical point. 
For $\q=0$ one finds
\bea
R^0_{\q=0}(t,s) &=& \left( \frac{s+\tm}{t+\tm}\right)^{3/2}\,,
\label{Rgauxeq0}\\
C^0_{\q=0}(t,s) &=&  \frac{1}{2}\frac{(s+\tm)^4-\tm^4}{[(s+\tm)(t+\tm)]^{3/2}}\,.\label{Cgauxeq0}
\eea
Comparing this results with $\tm = 0$, 
with the scaling forms~\reff{scalR2i} and~\reff{scalC2i}
we can identify $z=2$, $a=0$, $\bt=-3/2$ [see Eq.~\reff{thetabar}] 
in agreement with standard 
mean-field exponents ($\delta=3$, $\nu=\beta=1/2$, and $\eta=0$)
and determine $\bar A_R = 1$, $\bar A_C = \frac{1}{2}$, $\bar \F_R(x)=1$,
and $\bar \F_C(x)=1$. 
Taking into account the Gaussian expressions for the response and correlation
function for $m_0=0$ (see, \eg\ Ref.~\cite{cg-rev}), one finds $A_R =
1$ and $A_C = 2$, and therefore ${\cal R}_R =1$ whereas ${\cal R}_C =
\frac{1}{4}$. 
%
%

\begin{figure}[t]
\begin{center}
\epsfig{width=16truecm,file=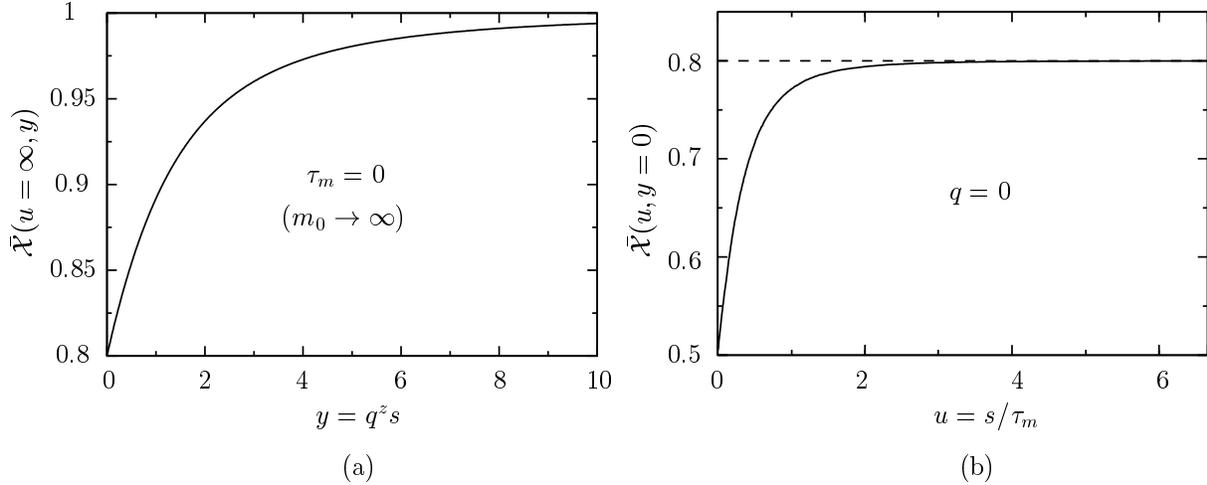}
\caption{%
Scaling function of the Gaussian FDR 
${\mathcal X}_\q(t,s) = \bar {\mathcal X}(u=s/\tau_m,y=q^zs)$ 
for (a) $m_0\rightarrow\infty$ as a function of $y$
(see also Fig.~1 of 
Ref. \protect\cite{gspr-05}) %
and (b) $\q=0$
and finite $m_0$ as a function of $u$.%
}
\label{FDRfig}
\end{center}
\end{figure}

%%% FDR
Let us consider in more detail the \fd ratio. Using Eq.~\reff{Rgaux}
one easily finds $\partial_t R^0_\q(t,t') = \delta(t-t') - [\q^2 + r_0
+ m_G^2(t)] R^0_\q (t,t')$, and therefore, from Eq.~\reff{Cgaux},
\bea
\partial_s C_\q^0(t,s) &=& 2 \int_0^\infty\rmd t'
R^0_\q(t,t')\partial_s R^0_\q(s,t') \nonumber\\
&=& 2  R_\q^0(t,s) -  [\q^2 +r_0 + m_G^2(s)] C_\q(t,s) \;.
\eea
Accordingly, at the critical point ($r_0=0$)
\be
{\cal X}_\q(t,s) = \frac{R^0_\q(t,s)}{\dpar_s C^0_\q(t,s)}
= \frac{1}{2} \left\{ 1 - [\q^2 + m^2(s)]\frac{C^0_\q(t,s)}{2
R^0_\q(t,s)}\right\}^{-1} \equiv 
\bar{\cal X}(s/\tau_m,q^2 s)\,,
\ee
which, as one can see from Eqs.~\reff{Rgauxe} and~\reff{Cgauxe}, 
actually depends on the scaling variables $u = s/\tau_m = 2 s m_0^2/3$ and 
$y = |\q|^z s = q^2 s$ whereas it is independent
of $t$, a typical property of the Gaussian approximation. 
The function
$\bar {\cal X}(u=0,y)$, corresponding to the case of a quench from
an high-temperature state, has already been investigated in
Ref.~\cite{cg-02a1}: For $y=0$ one has $\bar{\cal X}(0,0) =
\frac{1}{2}$, whereas it monotonically increases towards the
asymptotic value $\bar{\cal X}(0,y\rightarrow\infty) = 1$ as $y$
increases. This suggests that the violation of the FDT for long times 
$s\rightarrow \infty$ comes only from the
zero mode $\q=0$, whereas all the other modes properly
equilibrate. 
Let us consider the case $m_0\rightarrow\infty$, \ie\ $u\rightarrow \infty$.
From Eqs.~\reff{Rgauxe} and~\reff{Cgauxe} one finds 
\be
\bar{\cal X}(u=\infty,y) =
\frac{8 y^4}{9 -12 y +6 y^2 +8 y^4-
3 \rme^{-2 y}(3 + 2 y)}\,,
\label{Xqt}
\ee 
which monotonically increases from the value $\frac{4}{5}$ at $y=0$ to
$1$ for $y\rightarrow\infty$ [see Fig.~\ref{FDRfig}(a)]. 
As in the previous
case, the \fd theorem is satisfied for long times by
fluctuations with $\q\neq 0$, whereas it is violated by the zero mode
$\q=0$ (\ie\ by the global order parameter of the system), 
for which the FDR takes the value $\frac{4}{5}$.
As anticipated, this behavior coincides with that one found from the
exact solution of the dynamics of
the $d$-dimensional Ising model with
Glauber dynamics, in the limit of large $d$
[cp.~Eq.~(47) of Ref.~\cite{gspr-05} with $w \mapsto y$ to
Eq.~\reff{Xqt}].

Let us now focus on the FDR for the zero mode $\q=0$ (\ie\ $y=0$), as a 
function of $u=s/\tau_m$, given by [see Eqs.~\reff{Rgauxeq0} 
and~\reff{Cgauxeq0}]
\be
\bar{\cal X}(u,y=0) = \frac{4}{5}\left[ 1 + \frac{3}{5}
\left(1 + u\right)^{-4}\right]^{-1}\;.
\label{Xq0mgen}
\ee
This quantity increases monotonically from the value $\frac{1}{2}$ for
$u=0$, to the value $\frac{4}{5}$ for $u\rightarrow\infty$. This
implies that the FDR for the global order parameter (the total magnetization
in magnetic systems) in the long-time limit approaches $\frac{4}{5}$ whenever
the mean value of the initial magnetization $m_0$ is non zero (and the value
is indeed independent of the actual $m_0\neq 0$), as is the case
for quenches from the ordered state,
whereas the FDR is equal to $\frac{1}{2}$ for $m_0=0$, \ie\ for quenches from
a disordered initial state. 
The plot of ${\cal X}_{\q=0}(s,t)$ as a function
of $u = s/\tau_m$ is reported in Fig.~\ref{FDRfig}(b).

In the case of a quench from the disordered phase it was instructive to 
look at the FDR in the real space ${\bf x}$ which has the same
asymptotic value as ${\mathcal X}_\q(t,s)$ but a slower approach to 
it~\cite{cg-rev}. 
The expressions for the space-dependent Gaussian response
and correlation functions can be worked out by Fourier transforming
the corresponding equations~\reff{Rgaux} and~\reff{Cgaux}. We do not
report the result for the general case but we focus instead on the
expressions for $\x =0$, \ie\ on the autoresponse and autocorrelation
functions ($t>s$):
\bea
R^0_A(t,s) &=& \int(\rmd q) R^0_\q(t,s) =  K_d (t-s)^{-\dm} \left(\frac{s+\tm}{t+\tm}\right)^\frac{3}{2}\,,\label{RA}\\
C^0_A(t,s) &=& \int(\rmd q) C^0_\q(t,s) = 2 K_d
[(t+\tm)(s+\tm)]^{-\frac{3}{2}}\int_0^s\rmd t' \, (t+s-2t')^{-\dm} (t'+\tm)^3\,,
\eea
where $K_d = (4\pi)^{-\dm}$, and the expression~\reff{mgauss} has been
used. 
Introducing the scaling variables $x=s/t$ and $u = s/\tm$, and the function 
$I_a(x,u) = \int_0^1 \rmd v [1 + x(1-2 v))]^{-a}(1 + u v)^3$, 
one finds
\be
\frac{\partial_s C^0_A(t,s)}{R^0_A(t,s)} - 2 = - \frac{(1-x)^\dm}{(1 +
u)^3} \left[3 \frac{u}{1+u} I_\dm(x,u) + d
\,x  I_{\dm+1}(x,u)\right]\,.
\label{fdrA}
\ee
Note that $I_a(0,u) = [(1+u)^4-1]/(4 u)$ and
therefore the previous expression becomes, for $x=0$,
\be
\left.\frac{\partial_s C^0_A(t,s)}{R^0_A(t,s)}\right|_{x=0} - 2 =
-\frac{3}{4} \left[ 1 - (1+u)^{-4}\right] \;.
\label{XA000}
\ee
A comparison with Eq.~\reff{Xq0mgen} 
shows that
\be
\left. \frac{\partial_s C^0_A(t,s)}{R^0_A(t,s)}\right|_{x=0} = 
\left. \frac{\partial_s C^0_\q(t,s)}{R^0_\q(t,s)}\right|_{\q=0}\;,
\ee
\ie\ in the limit $t\rightarrow \infty$ the FDR computed from the
observables at $\x=0$ (local) is the same as that one computed from
the observables at $\q=0$ (global), for every value of $u = s/\tau_m$. 
This agrees with the heuristic argument of  Ref.~\cite{cg-02a1}, which
can be extended also to the case of non-vanishing initial
magnetization.  This has been explicitly checked in the case of infinite 
dimensionality in Ref. \cite{gspr-05} .

Consider now the case $x = s/t \rightarrow 1$, \ie\ the
regime characterized by a short time difference 
$\delta t \equiv t-s \ll s \rightarrow \infty$.
One expects on general grounds~\cite{cg-rev,cdk-97} 
that the system is in quasi-equilibrium with 
$R^0_A(t,s)=\partial_s C^0_A(t,s)$, as one verifies 
explicitly: Indeed the correlation function can be written in terms of
the scaling variables $x$ and $u$ as
\be
C_A^0(t,s) = 2 K_d t^{-\dm+1}
\left[\left(1+\frac{u}{x}\right)(1+u)
\right]^{-\frac{3}{2}} x I_\dm(x,u)\,.
\ee
Note that for $a >1$, 
$I_a(x,u)= [2(a-1)]^{-1}(1+u)^3 (1 - x)^{-a+1} [1 +O(1-x)]$,
and therefore
\be
C_A^0(t,s) = \frac{K_d}{\dm-1} (t-s)^{-\dm+1} [1 + O(1-s/t)]\,
\ee
which is the equilibrium form of the two-point correlation function at
the critical point $C_A^{0{\rm (eq)}}(t-s)$ [we recall that $C^{0{\rm
(eq)}}_\q(t-s) = \rme^{-\q^2(t-s)}/\q^2$]. From Eq.~\reff{RA} one
easily finds $R_A^0(t,s) = R_A^{0{\rm (eq)}}(t-s) [1 + O(1-s/t)]$.
Note that these conclusions, leading to $R^0_A(t,s)=\partial_s
C^0_A(t,s)$, are independent of the actual value 
of $u = s/\tau_m$ and therefore of $m_0$. 

In the following we shall consider the thermoremanent magnetization
defined as $\rho(t,s) = \int_0^s\rmd t'\,R_A(t,t')$. 
According to Eq.~\reff{RA} it can be written in the scaling form
\be
\rho^0(t,s) = K_d t^{-\dm + 1}
\left(1+\frac{u}{x}\right)^{-\frac{3}{2}} x J_\dm(x,u)
\ee 
where $J_a(x,u) = \int_0^1\rmd v\, (1 - v x)^{-a}(1+ v
u)^\frac{3}{2}$, with $J_a(0,u) = \frac{2}{5}[(1+u)^\frac{5}{2}-1]/u$
and $J_a(x,u) = (a-1)^{-1}(1+u)^\frac{3}{2}(1-x)^{-a+1}[1+O(1-x)]$.
Therefore within the Gaussian approximation one finds:
\be
\frac{\rho^0(t,s)}{C_A^0(t,s)} = \frac{1}{2} (1+u)^\frac{3}{2} \frac{J_\dm(x,u)}{I_\dm(x,u)}\;.
\ee
which gives, for $x=0$
\be
\left. \frac{\rho^0(t,s)}{C_A^0(t,s)} \right|_{x=0} = \frac{4}{5}
(1+u)^\frac{3}{2} \frac{(1+u)^\frac{5}{2} -1}{(1+u)^4-1} \, .
\ee
Note that, although for generic values of $u$ this expression differs
from $R^0_A(t,s)/\partial_sC_A^0(t,s)$ given in Eq.~\reff{XA000}, they
have the same asymptotic values for $u=0$ and $u\rightarrow
\infty$. On the basis of scaling forms it is easy to realize that this
equality holds beyond the Gaussian model and indeed in
Sec.~\ref{sec-mc} we shall compute the limiting FDR via the ratio
between the thermoremanent magnetization $\rho(t,s)$ and the
correlation function $C_A(t,s)$. 
\begin{figure}[t]
\begin{center}
\epsfig{width=10truecm,file=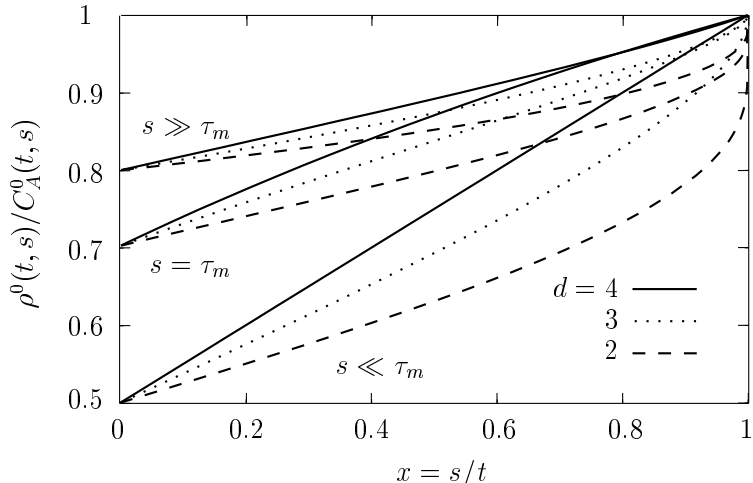}
\caption{Scaling function of $\rho(t,s)/C_A(t,s)$ as a function of the
scaling variables $x=s/t$ for fixed $u=s/\tau_m$, within the Gaussian
approximation for various values of the dimensionality $d$. (The
Gaussian approximation becomes exact for $d>4$.) 
From top to bottom $u=\infty$, $u=1$, and $u=0$. Some of the
qualitative features displayed by the Gaussian approximation are also
found in the numerical results presented in Sec.~\ref{sec-mc} (cp. the
plot for $d=2$ to Fig.~\ref{x_oo}(a)).}
\label{FDRAfig}
\end{center}
\end{figure}
In Fig.~\ref{FDRAfig} the ratio $\rho^0(t,s)/C_A^0(t,s)$ is shown for
fixed values of $u = s/\tau_m$ (from top to bottom: $u=\infty$, $u=1$,
$u=0$) as a function of $x = s/t$ and of the
dimensionality $d$. In spite of the fact that 
the Gaussian approximation provides exact
results only for $d>4$ it displays correctly some
qualitative features also for $d<4$, although reliable quantitative
predictions can be obtained only after having accounted for
fluctuations. In particular Fig.~\ref{FDRAfig} shows that
$\rho^0(t,s)/C_A^0(t,s)$ is an increasing function of $x$ for fixed
$u$, with an increasing concavity upon decreasing the dimensionality
$d$. In particular the function is well approximated by a straight
line followed by a sudden increase towards the
value $1$, for $x\rightarrow 1$. This behavior is more pronounced
upon decreasing $d$.
The limiting value for $x\rightarrow 1$ is expected to be $1$
also beyond the Gaussian approximation, because it is a signature of the
quasi-equilibrium regime, whereas the asymptotic value
for $x\rightarrow 0$ depends not only on $u$ (as correctly displayed
by $\rho^0(t,s)/C_A^0(t,s)$) but also on the dimensionality $d$ (we
shall verify in the next section for the case $u=\infty$).

\section{One-loop Fluctuation-Dissipation ratio}
\label{sec-onel}

In this section we present the one-loop (i.e., $O(\e)$ in the
$\e$-expansion where $\e = 4 - d$)
perturbative computation of the correlation and response functions for a 
quench from a state with initial magnetization 
$m_0\to \infty$ to the critical point. As we have 
argued from general scaling arguments and explicitly  shown within the 
Gaussian approximation, a finite $m_0$ only gives corrections
to the leading long-time behavior. The first step in this 
derivation is the one-loop equation of motion  obtained in 
Refs.~\cite{bj-76,bej-79} which we rederive in the next subsection for
the sake of 
completeness. Then we calculate the connected correlation and response 
functions in the 
two following subsections, reporting all the details of the
computation in Appendix \ref{appR} and \ref{appC}.
From these results 
we determine the FDR and the universal amplitude ratios ${\cal R}_R$
and ${\cal R}_C$.

\subsection{The equation of motion and its solution}

At one-loop level the tadpole contribution ${\cal T}$ to the 
equation of motion~\reff{eqmo} is
\bea
I(t)&=&\int (\rmd q) C^0_\q(t,t)=
2 \int_0^t \rmd t' \frac{t'^3}{t^3}\int (\rmd q)  \rme^{-2q^2(t-t')} =
%2 \int_0^t \rmd t'\frac{1}{[8\pi (t-t')]^{d/2}}\frac{t'^3}{t^3}=
\nonumber\\&=&
%2 t^{1-d/2} \int_0^1 \rmd x\frac{(1-x)^{-d/2} x^3}{(8\pi)^{d/2}}=\nonumber\\
t^{1-d/2} %\frac{192}{(8\pi)^{d/2}(8-d)(6-d)(4-d)(2-d)}
\frac{6 \Gamma(1-\dm)}{(8\pi)^\dm \Gamma(5-\dm)}
\equiv 2 N_d r_d  t^{1-d/2}\,,
\label{Idit}
\eea
where Eq.~\reff{Cgauxe} with $\tm=0$ has been used.
For later convenience we introduce $N_d=2/[(4\pi)^{d/2}
\Gamma(d/2)]$, 
\be
r_d=-\frac{3}{4\e} +\left[\frac{9}{16}-\frac{3}{8} (\gamma_E+\ln2)
\right]+O(\e)\equiv \frac{c_{-1}}{\e} +c_0+O(\e) \,,
\ee
and $\gt_0=N_d g_0$. Therefore Eq.~\reff{eqmo} becomes
\be
0=\dpar_t m +\frac{1}{3} m^3+\gt_0 r_d t^{1-d/2} m\,.
\ee
We look for a solution of the form $m(t)=m_G(t)[1+\gt_0 \rho(t)] +
O(\gt_0^2)$,  where $m_G^2(t)=3/(2t)$ 
is the Gaussian solution. It is straightforward to show that 
$\rho(t)=r_d/(3-d/2) t^{\e/2}$, so that
\be
m(t)=\sqrt{\frac{3}{2t}}\left[1-\gt_0 \frac{r_d}{3-d/2} t^{\e/2}\right]
+ O(\gt_0^2)\,.
\label{m1loop0}
\ee
Expanding in $\e=4-d$, we obtain the known result \cite{bj-76,bej-79}
\bea
m(t)&=&%\sqrt{\frac{3}{2t}} 
%\left[1+\gt_0 \frac{1}{1+\e/2}(c_{-1}/\e+c_0)(1+\e/2 \ln t)\right]=\nonumber\\
%&&
 \left(1-\gt_0 \frac{c_{-1}}{\e}\right) \sqrt{\frac{3}{2t}}
\left[1-\gt_0 \left(\frac{c_{-1}}{2} \ln t+c_0-\frac{c_{-1}}{2}\right)
+ O(\e^2,\e\gt_0,\gt_0^2)\right] \,.
\label{m1loop}
\eea
As expected, a dimensional pole $\sim 1/\e$ appears in the
expansion. To get finite renormalized quantities $(M,\gt)$ in the limit
$\e \rightarrow 0$ one has to renormalize the bare quantities
$(M_0,\gt_0)$ according
to the well-known renormalization procedure (see, \eg\ Ref.~\cite{zj}): 
$M_0 = Z^{1/2} M$, $\gt_0 = Z_g Z^{-2}\mu^\e \gt$ where $\mu$ is an
arbitrary momentum scale that we will neglect in the following,
$Z = 1 + O(\gt^2)$, and $Z_g = 1 + 3 \gt/(2\e) + O(\gt^2)$. Taking
into account Eq.~\reff{mresc} one recognizes in Eq.~\reff{m1loop} the
renormalization constant $Z_g^{1/2}/Z$. Once the renormalized $m(t)$
is expressed in terms of $\gt$, its scaling behavior at the RG fixed
point $\gt^* = 2\e/3 + O(\e^2)$ [for $d>4$, $\gt^*=0$ and the
predictions of the Gaussian model become exact] 
clearly shows up: $m(t)\sim
t^{-\varsigma} + O(\e^2)$ with 
$\varsigma =1/2(1+c_{-1}\gt^*) + O(\e^2) =1/2(1-\e/2)+O(\e^2) =
\beta/(\nu z)$~\cite{fr-76} [we recall that $z = 2 + O(\e^2)$,
$\beta/\nu = 1 - \e/2 + \eta/2 = 1 - \e/2 + O(\e^2)$].

\subsection{The response function}

\begin{figure}[tb]
\centerline{\epsfig{width=9truecm,file=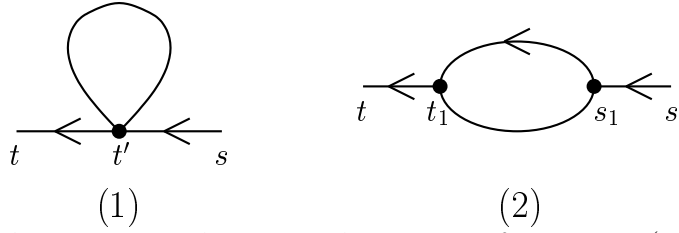}}
\caption{One-loop diagrams contributing to the response function
$R_\q(t,s)$. Directed lines represent response propagators
$R^0_\q(t,t')$, whereas undirected ones represent correlations
$C^0_\q(t,t')$. Causality implies that both diagrams vanish for $s>t$
(as a result of the directed line joining the external legs of the
diagrams). The corresponding expressions (reported in Appendix~\ref{appR}) 
are indicated with $I_1(t,s)$ and $I_2(t,s)$, respectively.}
\label{diagR}
\end{figure}

At one-loop order the expression~\reff{Rgauxe} for the response
function~\reff{Rgaux} gets modified because of the one-loop term
contributing to the magnetization $m(t)$, computed in
Eq.~\reff{m1loop}. Accordingly, $m^2(t) = 3/(2 t) - 3 \gt_0
r_dt^{1-\dm}/(3-\dm) + O(\gt^2)$, and therefore
\bea
R^0_\q(t,s) &=& \rme^{-\q^2(t-s) - \int_s^t\rmd t' m^2(t')}\nonumber\\
&=& \left(\frac{s}{t}\right)^{3/2} \left[ 1 + \gt_0 \frac{3}{3-\dm} r_d
\frac{t^{2-\dm}-s^{2-\dm}}{2-\dm} + O(\gt_0^2)\right]\rme^{-\q^2(t-s)}
\eea
In addition to this contribution two further terms come from the
interaction vertices of  ${\cal L}_1$ [see Eq.~\reff{L1}] and are
depicted in Fig.~\ref{diagR} [note that, up to this order, in computing the
diagrammatic contributions one can use the tree-level expressions
Eqs.~\reff{Rgauxe} and~\reff{Cgauxe} -- with $\tm=0$ -- for the
response and correlation function]. 
In terms of these diagrams the response function reads
\be
R_{\q=0}(t,s)= R^0_{\q=0}(t,s) -\frac{g_0}{2} I_1 + g_0 I_2  + O(g_0^2)\,.
\ee
The relevant integrals have been calculated in Appendix~\ref{appR},
Eqs.~\reff{I1} and~\reff{I2}. %
Collecting the three terms together we obtain, in terms of the
renormalized coupling $\gt$ ($x=s/t$)
\be
R_{\q=0}(t,s)= x^{3/2}
\left\{1+\gt
\left[-\frac{3}{8}\ln x+\frac{3}{4}
\left(\frac{\pi^2}{3} - \frac{7}{2} + f_R(x)
\right)\right]\right\} + O(\gt^2,\e\gt,\e^2)\,
\ee
where
\be
f_R(x) = 3 + \frac{x}{2} + 3 \frac{1-x}{x}\ln (1-x)-2 \Li_2(x) \,,
\ee
[we recall the definition of the dilogarithmic function $\Li_2(x)
\equiv \sum_{n=1}^\infty x^n/n^2$] is a monotonically increasing
function with $f_R(0) = 0$ and $f_R(1) = 7/2 - \pi^2/3$.
Note that all the dimensional poles and the non scaling terms
originally present in the expressions of $I_{1,2}$ cancel out in 
the sum, as they should.% 

At the fixed point $\gt^*=2\e/3 + O(\e^2)$ \cite{zj}, 
the response function can be written as Eq.~\reff{scalR2i}
%\be
%R_{q=0}= A_R \left(\frac{s}{t}\right)^\bt (t-s)^a F_R(s/t)
%\ee
with %(the normalization is $F_R(x)=1+O(x)$)
\bea
a&=&O(\e^2)\,,\\
\bt&=&-\frac{3}{2}+\frac{3}{8} \gt^* + O(\e^2) =-\frac{3}{2}+\frac{1}{4}\e+O(\e^2)\,,\\
\bar A_R&=& 1+\gt^*\left(\frac{\pi^2}{4}-\frac{21}{8}\right) + O(\e^2)=
1+\e\left(\frac{\pi^2}{6}-\frac{7}{4}\right)+O(\e^2)\,,
\label{AR}\\
\bar \F_R(x)&=& 1+ \frac{3}{4} \gt^* f_R(x) + O(\e^2) = 1+ \frac{\e}{2} f_R(x) + O(\e^2) \,.
\eea
$\bt$ agrees with the one-loop expression of $-\beta\delta/(\nu z)$ \cite{zj}.
Note that, as a difference with the case $m_0=0$ \cite{cg-02a1}, 
the response function present a one-loop correction to the Gaussian result.

In Ref.~\cite{cg-02a1} it was found that $A_R = 1 + O(\e^2)$,
leading to
\be
{\cal R}_R = \frac{\bar A_R}{A_R} = 
1+\e\left(\frac{\pi^2}{6}-\frac{7}{4}\right)+O(\e^2) \;.
\ee
A first numerical estimate of ${\cal R}_R$ in three and two spatial 
dimensions can be obtained
by setting $\e=1$ and $\e=2$ in the previous equation, giving 
${\cal R}_R (d=3) \simeq 0.9$ and ${\cal R}_R (d=2) \simeq 0.8$.
However, sizable systematic shifts of these estimates
might still come from higher-order terms in the
$\e$-expansion. 

\subsection{The connected correlation function}

\begin{figure}[th]
\centerline{\epsfig{width=12truecm,file=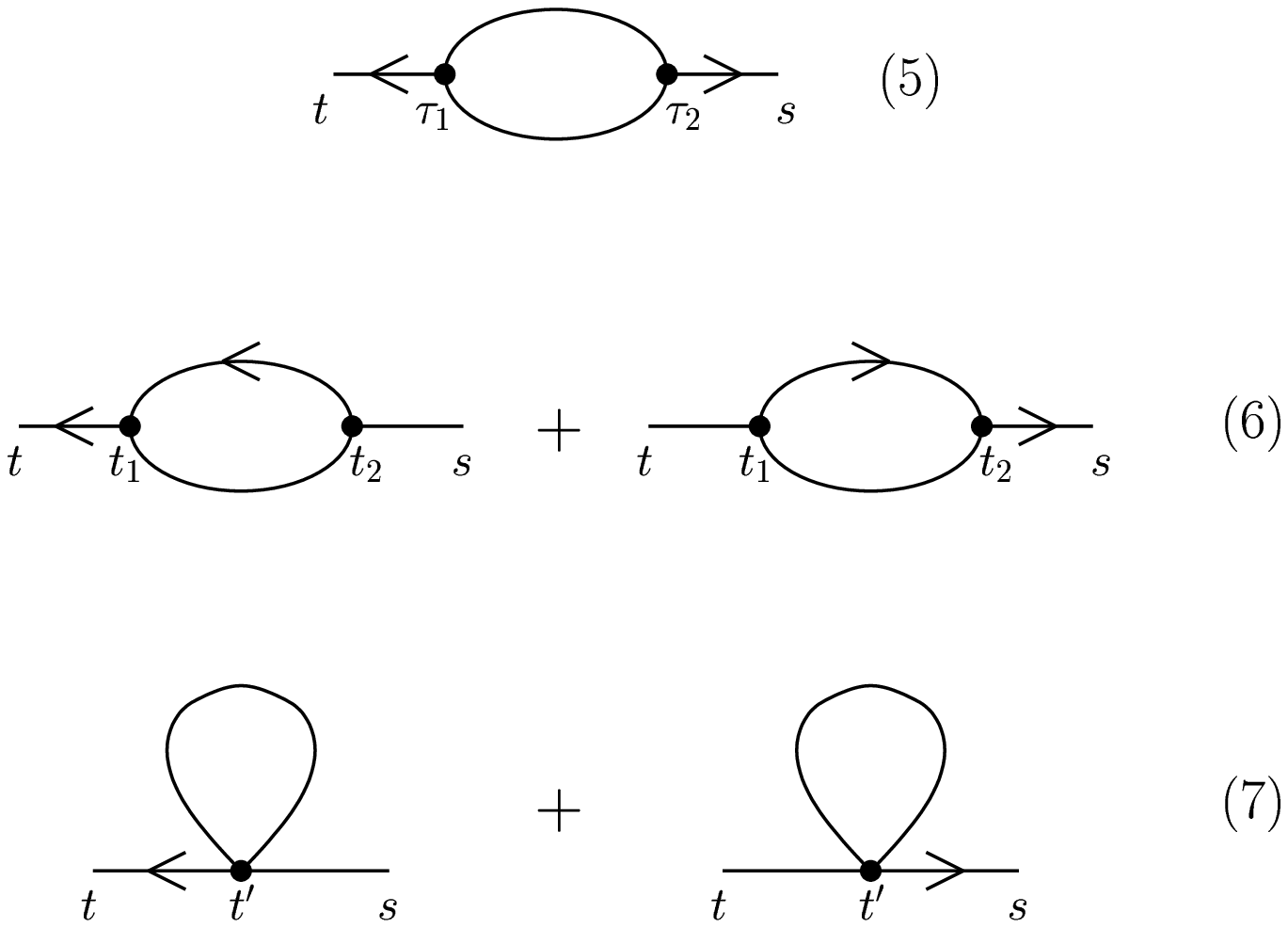}}
\caption{One-loop diagrams contributing to the correlation
function (see also the caption of Fig.~\ref{diagR}). The diagrams
represented here are invariant under the exchange $s\leftrightarrow
t$, and therefore one can assume $t>s$. Their corresponding
expressions are given, from top to bottom, 
by $I_5(t,s)$, $I_6(t,s)$, and $I_7(t,s)$ reported in
Appendix~\ref{appC}.}
\label{diagC}
\end{figure}

In terms of the Feynman diagrams depicted in Fig.~\ref{diagC}, the 
connected correlation function at one-loop reads
\be
C_{\q=0}(t,s) = C^0_{\q=0}(t,s) +\frac{g_0}{2} I_5+g_0 I_6-\frac{g_0}{2} I_7 + O(g_0^2)\,.
\label{Csum}
\ee

Using the results reported in Appendix~\ref{appC} and summing all the 
contributions we obtain the one-loop correlation function
\be
C_{\q=0}(t,s)=\frac{1}{2}s \,x^{3/2}
\left[1+\gt \left(-\frac{719}{320}+\frac{37}{160}\pi^2-\frac{3}{8}\ln x
\right)+\gt f_C(x) +O(\gt^2,\e\gt,\e^2)\right] \,,
\label{Cq0}
\ee
where $f_C(x)$ is given by Eq. (\ref{fx}) with $f_C(0)=0$.
Eq.~(\ref{Cq0}) can be cast in the scaling form (\ref{scalC2i}), 
with $\bt=-\beta\delta/(\nu z)$, $a=O(\e^2)$,
\be
2\bar A_C=1+\gt^* \left(-\frac{719}{320}+\frac{37}{160}\pi^2\right)+O(\e^2)=
1+\e\left(\frac{37}{240}\pi^2 -\frac{719}{480}\right)+O(\e^2)\,,
\label{AC}
\ee
and 
\be
\bar \F_C(x)=1+ \frac{2\e}{3} f_C(x) + O(\e^2) 
\ee

Taking into account that, up to one loop, $A_C = 2 (1 +\e/6) +
O(\e^2)$~\cite{cg-02a1}, one finds 
\be
{\cal R}_C = \frac{\bar A_C}{A_C} = \frac{1}{4} \left[1 + \e \left(\frac{37}{240}\pi^2
-\frac{799}{480} \right) \right] + O(\e^2)\,,
\ee
leading to the numerical estimates ${\cal R}_C(d=3) \simeq 0.21$ 
and ${\cal R}_C(d=2) \simeq 0.18$.

\subsection{One-loop FDR}
\label{subsec-1loopFDR}

It is now easy to compute from Eqs.~(\ref{AR}) and (\ref{AC}) the FDR
\be
X^\infty=\frac{\bar A_R}{\bar A_C(1-\bar \theta)}=
\frac{4}{5}-\left(\frac{73}{600}-\frac{\pi^2}{100}\right)\e+O(\e^2)\,.
\label{Xinfeps}
\ee
Let us comment on the main features of $X^\infty$.
Close to $d=4$ it decreases as the dimensionality decreases.
Probably a monotonically decreasing behavior has to be expected down to 
$d=1$ (the lower critical dimension) where it is known that 
$X^\infty=1/2$ \cite{ms-04}.
One can provide  numerical 
estimates of $X^\infty$ in physical dimensions $d=3$ and $d=2$ 
by evaluating  Eq.~(\ref{Xinfeps}) for $\e=1$ and $\e=2$,
respectively, leading to $X^\infty(d=3) \simeq 0.78$ and 
$X^\infty(d=2) \simeq 0.75$. 
These values are mainly indications of the actual 
ones, since higher-order terms in the $\e$-expansion  
might give relevant contributions.
However, the smallness of the one-loop correction (which is just the
$3\%$ and $6\%$ in $d=3$ and $2$, respectively, of 
the mean-field result) strongly suggests that such estimates can work 
effectively, despite the low order considered in perturbation theory. 
For this reason in the next section we proceed to a numerical 
evaluation of $X^\infty$ by means of an extensive Monte Carlo simulation of 
the two-dimensional Ising model with Glauber dynamics.
%%%%
%%%%
In Ref.~\cite{gspr-05} a phenomenological evolution equation for the
fluctuating magnetization $\langle \p(\q=0,t)\rangle$ [corresponding
to a tree-level approximation of the actual Langevin equation~\reff{lang}] is
used [see Eq.~(52) therein] to explain the values of $X^\infty$ found
within the Gaussian approximation. The resulting expression 
$X^\infty = (2\beta + 3\nu z)/(2\beta+4\nu z)$ for $X^\infty(m_0\neq
0)$
%, 
can be naively used 
beyond the original approximation, giving 
$X^\infty = (2\beta + 3\nu z)/(2\beta+4\nu z) = 4/5 - \e/50 +
O(\e^2)$ which provides a rather good estimate of the actual
result Eq.~\reff{Xinfeps}.

\section{Monte Carlo simulations}
\label{sec-mc}

Monte Carlo (MC) 
simulations have been widely used to investigate the problem of
the non-equilibrium critical dynamics and aging 
in two dimensions (see for instance
\cite{gl-00i,mbgs-03,sdc-03,ch-03,frt-03,ch-04,ak,gkf-03,f-05,barrat,clz-04,pg-04} 
and references
therein) providing 
results which are in rather good agreement with those obtained within
different analytical approaches (see, \eg\ Ref.~\cite{cg-rev}).
It is then natural to test our predictions 
on the critical aging from a magnetized state with short-range
correlations by comparing them with 
MC simulations of two-dimensional Ising model defined by the 
Hamiltonian (\ref{HON}) and with spin-flip dynamics.

\subsection{Details of the simulations}
We performed our Monte Carlo simulations using the Heat Bath updating rule,
simulating a large system of $N=L\times L$ (with $L=10^3$ and periodic
boundary conditions)
spins $S_i$, $i=1,\ldots,N$ and averaged our
observables over more than $3000$ different runs.  
We computed the magnetization
\be
M(t) = \left< \frac{1}{N} \sum_{i=1}^N S_i(t) \right> \,,
\ee
the
{\it connected} two-time (auto)correlation function
\be C_A(t_1,t_2) =\left< \frac{1}{N} \sum_{i=1}^N S_i(t_1) S_i(t_2)
  \right> - M(t_1) M(t_2) \,, 
\ee
where $\left< \ldots \right>$ 
stands for the MC average. We also considered
the thermoremanent magnetization  (TRM) at a site $i$ and time $t_2$
for a field applied at the same site from time $t=0$ to $t_1$:
\be 
\rho (t_1,t_2) = \int_0^{t_1} \rmd t' R_A(t_2,t') , 
\ee
where $R_A(t,s)$ is the (auto)response function of the system at time
$t$ and site $i$ to a magnetic field 
$h_i$ applied at the same site $i$ but at an earlier time $s$. 
In MC simulations a
random probing field $h_i$ (with $\overline{h_i}=0$ and $\overline{h_i h_j} =
h^2\,\delta_{i,j}$) is usually used to extract the response function. However,
measuring numerically 
the response in the case of a uniformly magnetized system is
harder than in the usual case where $M(t)=0$. Indeed one
usually applies a
small field $h_i$ at lattice site $i$, checks that the system remains 
in the linear regime, measures
the local magnetization $M_i\equiv \langle S_i\rangle$ at a later time 
and finally deduces the response via
$M_i/h_i$~\cite{barrat}. Here, however, one would need to compute
$[M_i(h)-M_i(0)]/h_i$ which involves two different precise measures of the
magnetization in the presence {\it and} in the 
absence of the field $h_i$. Instead of
using this numerically demanding procedure, we follow the approach
proposed in Refs.~\cite{ch-03,frt-03}, and later used in 
Ref.~\cite{gkf-03,f-05,ff-05} which
allows for the determination of the response function without introducing
any magnetic field in the simulation. This approach has been adopted
in Ref.~\cite{gkf-03}
to compute $X^\infty$ [see Eqs.~\reff{dx} and~\reff{xinfdef}] 
for the two-dimensional Ising model with conserved and
non-conserved dynamics leading to satisfactory results. For details on the
method, we refer the reader to Refs.~\cite{frt-03,gkf-03}. 
Here, we just recall the formula we used to compute the TRM:
\be 
T\rho (t_1,t_2) = \frac{1}{N} \sum_i \left < S_i(t_2) \Delta
  S_i(t_1)\right>,  
\ee
where the function $\Delta S_i(t_1)$ is computed during the simulation from
time $t=0$ to $t_1$ and is defined by
\be 
\Delta S_i(t_1) = \sum_{s=0}^{t_1} \delta_{{\cal I}(s),i} \left[ S_i(s) -
  \tanh(h^w_i(s)/T) \right]\,,  
\ee
where the function ${\cal I}(s)$ gives the index of the spin to be updated at
time $s$ (which depends on the random update performed in the MC
simulation), $T$ is the temperature ($T=T_c$ in what follows) 
and $h^w_i(s)$ is the Weiss local magnetic field accounting
for the effect of the interaction specified by the Hamiltonian~\reff{HON} 
on the spin to be updated at time $s$ (in the present
case it is just the sum of the value of the neighboring spins, \ie\
$h^w_i(s) = \sum_{j:|i-j|=1} S_j(s)$). Using this
method, the functions $C_A(t_1,t_2)$ and $\rho(t_1,t_2)$ can be calculated in
the same simulation for any value of $t_1$ and $t_2$, allowing
for very efficient simulations.
The initial state of each MC run is prepared such that 
$\langle M(t=0) \rangle_{\rm ic} = M_0$ ($|M_0|\le 1$) where 
$\langle\ldots\rangle_{\rm ic}$ stands for the average over the
different realizations of the initial condition. 
In particular the
value of each single spin in the lattice is chosen to be $+1$ and $-1$ with
probability $p_+ = (1+M_0)/2$ and $p_- = 1-p_+$, respectively, leading to 
$\Delta_0 \equiv \langle[M(t=0) - M_0]^2\rangle_{\rm ic} = (1 - M_0^2)/N$, and
vanishing initial correlations. Given that $\tau_0 \sim
\Delta_0^{-1}$  [see Eq.~\reff{idist}] one generally expects
corrections to the leading scaling behavior which could in principle
be reduced by preparing the initial state with {\it sharply fixed}
magnetization.

\subsection{Scaling forms}
\begin{figure}
  \epsfig{width=\textwidth,file=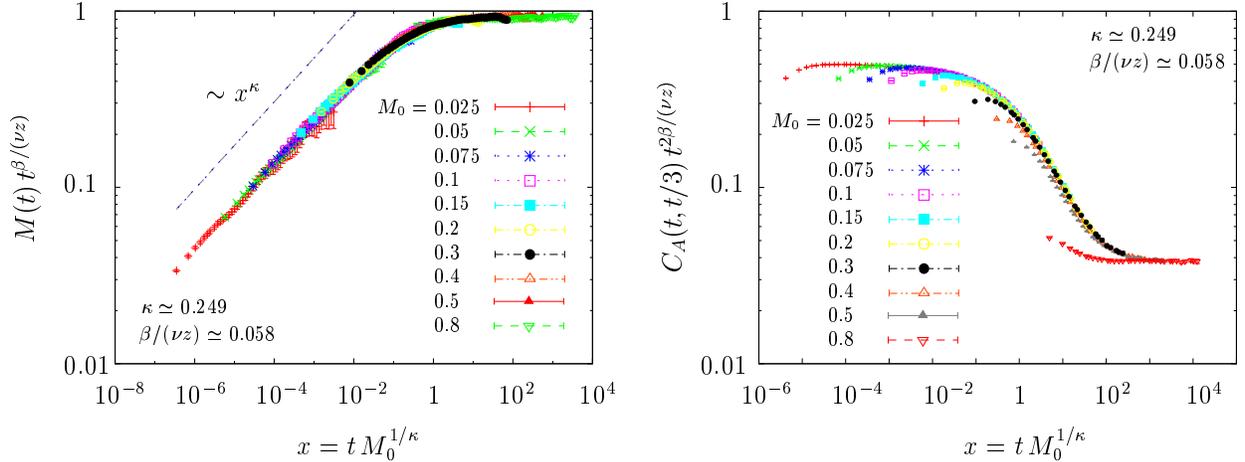}
\caption{Scaling of the magnetization (left) and of the connected correlation
  function (right) for different initial conditions with magnetizations $M_0$.}
\label{scaling_M}
\end{figure}

We now present the numerical results obtained using the method
described in the previous subsection.  
We first consider the scaling behavior 
of the magnetization and of the correlation as functions of 
the initial magnetization $M_0$. 
According to the theoretical prediction Eq.~(\ref{scalm0}), 
we expect the magnetization to scale as
\be 
M(t,M_0) = t^{-\frac{\beta}{\nu z}} \widehat F_M(t M_0^{1/\k}) = 
t^{-0.057} \widehat F_M(t M_0^{4.02})\,, 
\label{mc_1}
\ee
where we have used the value of the critical exponent for the
two-dimensional Ising model: $\beta =1/8$, $\nu=1$, $z=2.1667(5)$, and
$\theta=0.383(3)$ (see,\eg\ Ref.~\cite{cg-rev}), leading to $\k \simeq
0.249$ [see after Eq.~\reff{scalm0}] 
and $\beta/(\nu z) \simeq 0.058$. For the short-time behavior
one expects $\widehat F_M(x \rightarrow  0) \sim x^\k = x^{0.249}$. 
This scaling is rather well displayed by our data, as shown in 
Fig.~\ref{scaling_M}. 

Having tested the initial increase of the magnetization, we now consider 
the connected correlation function. 
In the previous sections we
focussed on the scaling properties of the
correlation for vanishing momentum $C_{\q=0}(t,s)$,
whereas now we are interested in the behavior of the
(auto)correlation function $C_A(t,s) = \int (\rmd q)
C_\q(t,s)$. According to the discussion in Sec.~\ref{sec-scaling-forms},
the scaling form of $C_\q(t,s)$ is given by Eq.~\reff{scalC} where
the scaling function has an additional dependence on the 
variable $y\sim q^z(t-s)$. Accordingly we expect
\be 
C_A(t,s) = s(s/t)^{\frac{\beta \delta}{\nu z}}
(t-s)^{a-\frac{d}{z}} \widehat F_C(s/t, M_0 t^\k) \,,
\label{mc_2}
\ee
where the change $a \mapsto a -d/z$  is due to the integration over
the momenta.
To make evident the scaling with respect to
$M_0$, we consider the correlation $C_A(t,t/3)$ for different initial
magnetizations $M_0$. Thus we expect a data collapse for $t$ large enough
according to
\be 
C_A(t,t/3) = t^{-2\frac{\beta}{\nu z}}\widehat G_C(t M_0^\k) = t^{-0.11}\widehat G_C(t M_0^{4.02}).  
\ee
where we used the relation $1+a-d/z = -2\beta/(\nu z)$.
The corresponding data are reported in Fig.~\ref{scaling_M}. 
Apart from a transient short-$t$ effect, we observe, again, 
a rather good data collapse. 

\subsection{Correlations, Responses and $X^\infty$}

\begin{figure}
  \epsfig{width=\textwidth,file=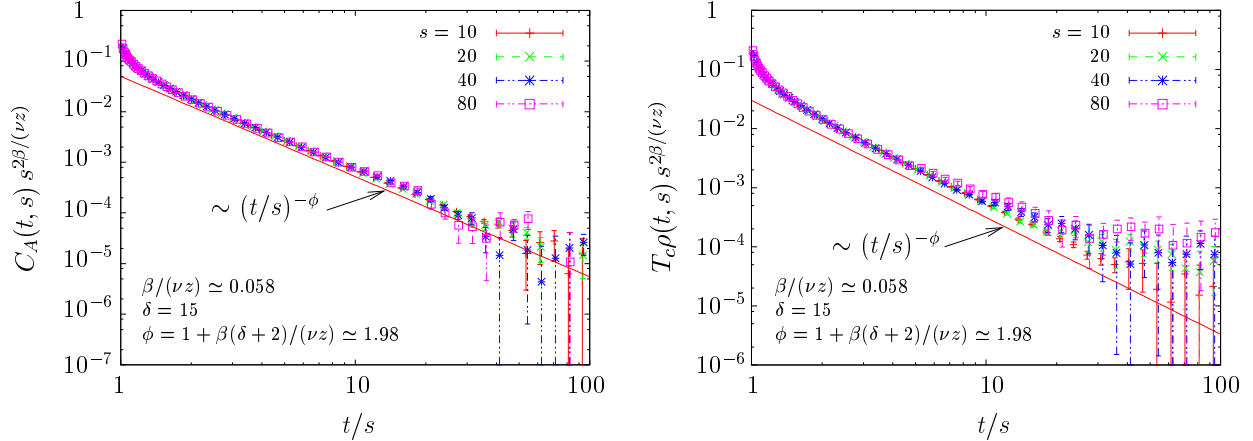}
\caption{Scaling of the connected two-times correlation (left) and integrated
  response (right) functions for completely ordered initial conditions using
  Eq.~(\ref{mc_3}).}
\label{scaling_CR}
\end{figure}

We now specialize to completely ordered initial condition ($M_0=1$). 
According to Eqs.~(\ref{scalR2i}) and (\ref{scalC2i}), the two-time connected
correlation function and the 
integrated response TRM are expected to share the same critical
scaling form which we write as
\be
 s^{2 \beta/\nu z}  C_A(t,s) = \widehat f_C(t/s)\,,\qquad
{\rm and}\qquad
 s^{2 \beta/\nu z}  \rho(t,s) = \widehat f_\rho(t/s)\,,
\label{mc_3}
\ee
with $2 \beta/\nu z \approx 0.116$ and $f_{C,\rho}(x\gg1) \sim x^{-\phi}$
where $\phi = -a+\frac{d}{z} + \frac{\beta\delta}{\nu z} =
1+\beta(\delta+2)/(\nu z)\simeq 1.98$ (in two
dimensions $\delta = 15$).
This behavior is again rather clearly displayed by our data, as
Fig.~\ref{scaling_CR} shows. Note however that due to the
fast decay of the correlations, it is difficult to obtain good data for
$x=t/s\gtrsim 20$ and that the TRM is affected by 
systematic deviations for the longer times.

\begin{figure}
\epsfig{width=\textwidth,file=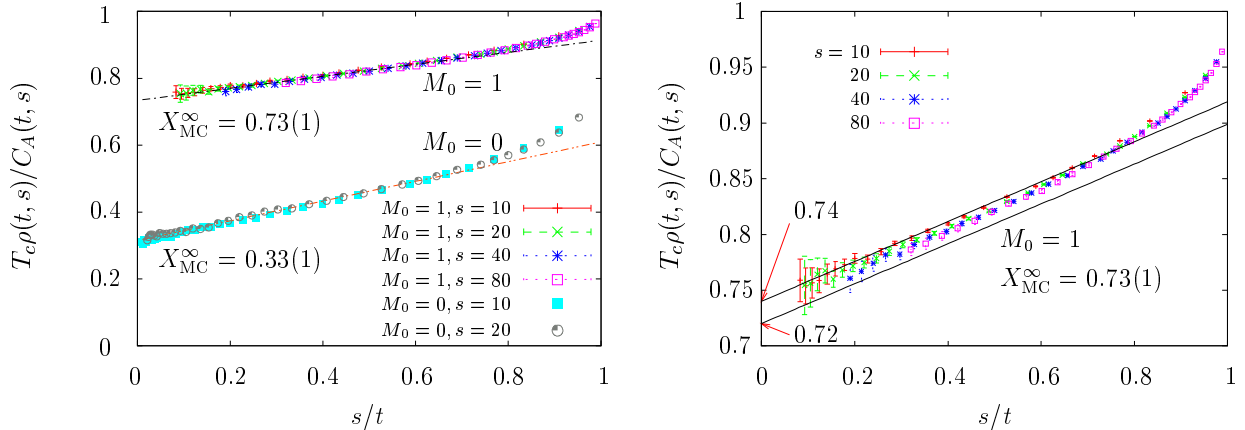}
\caption{$T_c{\rho(t,s)}/{C_A(t,s)}$ versus $s/t$ for disorder and 
  non disorder
  initial conditions (left). From our data we estimate 
$X^{\infty}_\MC=0.73(1)$ 
  for ordered initial conditions (right).}
\label{x_oo}
\end{figure}

Finally, we consider the value of the FDR that can be alternatively 
read off from the limit 
\be
X^{\infty} = \lim_{C_A \rightarrow 0} \frac{T_c \rho(t,s)}{C_A(t,s)} \,, 
\ee 
which indeed provides the estimate for $X^\infty$ as defined in
Eq.~\reff{xinfdef}.
The plot in Fig.~\ref{x_oo}  clearly shows
that $T_c \rho(t,s)/C_A(t,s)$ in Fig.~\ref{x_oo} is rather well  
approximated by a straight line in a wide range of $x=s/t$, both in
the case $M_0 =0$ and $M_0\neq 0$. (This point was already observed in
Ref.~\cite{mbgs-03}, see also \cite{comm}) 
The cross-over towards the quasi-equilibrium value $T_c
\rho(t,s)/C_A(t,s)=1$ takes place for $x\sim 0.75$--$0.85$ (although
the crossover is 
not complete in the set of data for the case $M_0=0$). A similar
qualitative behavior is displayed by the field-theoretical
Gaussian approximation for the same quantity, reported in
Fig.~\ref{FDRAfig}. A more quantitative comparison, however, would
require the analysis of the effects of fluctuations.
In order to provide the numerical estimate for $X^\infty$ we use
only the set of data which gives reasonable errors bars (typically
$t/s \lesssim 20$), reported in Fig.~\ref{x_oo}. From these data, we
roughly estimate 
\be
X^{\infty}_\MC = 0.73(1) \,, 
\label{XMC}
\ee
which compares rather well with the
one-loop prediction $X^\infty \simeq 0.75$. 
Using the
expression for $X^\infty$ provided in Ref.~\cite{gspr-05} (see also
Subsec.~\ref{subsec-1loopFDR}) as an improved Gaussian approximation 
one gets $X^\infty = [3 + 2\beta/(\nu z)]/[4 + 2\beta/(\nu z)] \simeq
0.757$ which is indeed rather close to the numerical result.

\section{Conclusions}
\label{sec-con}

In this paper we have considered the non-equilibrium dynamics of a system
in the Ising universality class starting from a state with initial 
magnetization $M_0$ and
evolving at criticality according to a purely dissipative dynamics. 
By means of RG arguments we found that in the aging regime
($t\gg s\gg \tau_m \sim M_0^{-1/\k}$) the zero-momentum 
response and {\it connected} 
correlation functions display the scaling behavior
\bea
R_{{\bf q}=0}(t,s) &=& \bar A_R\, (t-s)^a(t/s)^\bt \bar \F_R(s/t)\; ,\\
C_{{\bf q}=0}(t,s) &=& \bar A_C\,s(t-s)^a(t/s)^{\bt'} \bar \F_C (s/t)\; ,
\eea
independently of the actual value of $M_0\neq0$.
We showed that the exponents $\bt$ and $\bt'$ are not new non-equilibrium 
quantities, but they are given by standard equilibrium 
exponents as
\be
\bt = \bt' = - \frac{\beta\delta}{\nu z}\,.
\ee
In addition we showed that the ratio of the non-universal constants 
$\bar A_R$ and $\bar A_C$ with the corresponding ones obtained in the case 
$M_0=0$ are universal.

We solved exactly the dynamics within the Gaussian approximation and we derived
all the relevant universal ratios and scaling functions. 
Then we considered the effect of fluctuations in order to provide more
accurate predictions in physical dimensions $d=2,3$. 
We computed the first order correction
in $\e$-expansion to the response and to 
the {\it connected} correlation functions and 
from them we obtained the universal ratios and associated scaling functions. 
In particular, for the FDR we found 
\be
X^\infty=
\frac{4}{5}-\left(\frac{73}{600}-\frac{\pi^2}{100}\right)\e+O(\e^2)\,,
\ee
that gives $X^\infty\simeq 0.78$ in $d=3$ and $X^\infty\simeq 0.75$ in $d=2$.
On the other hand, if the FDR is defined from the {\it non-connected} 
correlation function then the asymptotic value would be zero because of
the scaling  in the volume and times of the additional 
term $M(t)M(s)$.

In order to confirm our theoretical predictions we
carried out extensive Monte Carlo simulation for the two-dimensional Ising
model with Glauber dynamics using the approach of 
Refs.~\cite{ch-03,frt-03}.  
We checked the predicted scaling forms for the two-time response
and correlation functions in addition to the well-known scaling
behavior of the magnetization~\cite{jss-89}.
Our analysis leads to a quite accurate determination of the FDR
\be
X^\infty_\MC=0.73(1)\,, 
\label{XMCconcl}
\ee
which is in rather good agreement with the field-theoretical
estimate. 

It is worth noticing that at criticality $X^\infty <1$ independently
of the initial magnetization. Therefore, insisting on the definition
of an effective temperature, one finds $T_{\rm eff} \equiv T_c/X^\infty
>T_c$. This fact can be naively understood for the case of a quench
from high temperature to the critical point: The system,
because of its slow dynamics, is somehow unable to transfer properly
heat to the thermal bath, resulting in an effective higher value of 
the temperature. In the case of a
sudden heating from a low-temperature configuration (with
non-vanishing $M_0$ and given initial short-range correlations
$\Delta_0$) to the critical point one would have expected 
(as observed in some cases~\cite{ff-05}) $T_{\rm eff}$ being larger
than the initial temperature but smaller than $T_c$. 
This is not the case here, showing that the
properties of $T_{{\rm eff}}$ at criticality are both non trivial and
counter-intuitive. 

Let us now compare our results with those available in the
literature. 
To our knowledge the ageing properties of the 
non-equilibrium critical relaxation from an ordered
(alternatively magnetized) state has been the subject of 
three recent studies~\cite{ft-05,gspr-05,as-05}.  
In Ref.~\cite{gspr-05} the lattice Ising model with Glauber dynamics
has been solved in the mean-field cases of (a) fully connected
lattice (infinite interaction range) and (b) nearest-neighbor
interaction in the limit of large dimensionality $d$. The
corresponding results are
completely equivalent to our Gaussian solution, as expected.  
In Ref.~\cite{as-05} the ageing properties of the spherical model with
relaxational dynamics have been considered. The scaling forms for
response and correlation functions found there agree with our general
predictions.  
In this case $X^\infty$ is a monotonic
function of the dimensionality $d$, 
interpolating between $1/2$ at $d=2$ and $4/5$ at
$d=4$.  This shares with our result for $d<4$ a downward correction to
the Gaussian result $4/5$. 
However a more quantitative comparison is clearly not
possible, since the two models belong to different universality
classes.  
The non-linear $\sigma$-model has been used to study the 
critical aging dynamics of ferromagnetic systems with 
$O(N)$ symmetry ($N>2$) in Ref.~\cite{ft-05}. Correlation and response
functions are computed in a dimensional $\tilde\e$-expansion around the lower
critical dimension $d=2$, with $\tilde\e=d-2$, leading to
$X^\infty=1/2+O(\tilde \e)$.  
These results do not compare directly with ours, since the
behavior of  $O(N)$ models is affected by the massless fluctuation modes in
the directions transverse to the decaying magnetization, in addition
to the longitudinal one.
In order to make a contact with Refs.~\cite{ft-05,as-05} we have also
studied the non-equilibrium aging dynamics of the $O(N)$ model up to
first-order in the $\e$-expansion. 
The analysis will appear elsewhere~\cite{cg-prep}.

\acknowledgments

We thank H.~W.~Diehl, M. Henkel, G.~Schehr, M. Pleimling, B.~Schmittmann, 
P. Sollich, and U.~T\"auber for very useful discussions. 
PC acknowledges the financial support from the 
Stichting voor Fundamenteel Onderzoek der Materie (FOM).

\appendix

\section{Feynman diagrams for the response function}
\label{appR}

There are two one-particle irreducible (1PI, \ie\ that cannot be
disconnected by cutting one propagator) diagrams
entering in the calculation of the response
function [see Fig.~\ref{diagR}]. One is $I(t)$ [see Eq.~\reff{Idit},
depicted in Fig.~\ref{diagR}(1)] and the other is the ``bubble'' in Fig.~\ref{diagR}(2):
\be 
B_{RC}(t_1,s_1)=\int (\rmd q) R^0_\q(t_1,s_1) C^0_\q(t_1,s_1) \,.  
\ee 
For $t_1 > s_1$ [because of causality, $B_{RC}(t_1\le s_1,s_1)=0$]
\bea
B_{RC}(t_1,s_1)&=&\int (\rmd q) \left(\frac{s_1}{t_1}\right)^{3/2} \rme^{-\q^2(t_1-s_1)}
\frac{2}{(t_1s_1)^{3/2}}\rme^{-\q^2(t_1+s_1)}\int_0^{s_1} \rmd x x^3 \rme^{2\q^2x}\\
%&=& \frac{2 t_1^{-3}}{(4\pi)^\dm}\int_0^{s_1} d x x^3 (2(t_1-x))^{-\dm}\\
&=& 2(8\pi)^{-\dm} H_d(t_1,s_1)
\eea
with
\bea
H_d(t_1,s_1)&=&t_1^{-3} \int_0^{s_1} \rmd x\; x^3 (t_1-x)^{-\dm}
\nonumber\\&=&
A_d t_1^{-1+\e/2}-A_d t_1^{-1}(t_1-s_1)^{\e/2}+B_d t_1^{-2}s_1 (t_1-s_1)^{\e/2}
+C_d t_1^{-3}s_1^2 (t_1-s_1)^{\e/2} \nonumber\\&&
+D_d t_1^{-3}s_1^3 (t_1-s_1)^{-1+\e/2}\,,
\label{para}
\eea
where
\bea
A_d&=&\frac{6}{(2+\e/2)(1+\e/2)\e/2(-1+\e/2)}\,,\\
B_d&=&-\frac{6}{(2+\e/2)(1+\e/2)(-1+\e/2)}\,,\\
C_d&=&-\frac{3}{(2+\e/2)(-1+\e/2)}\,,\\
D_d&=&-\frac{1}{(-1+\e/2)}\,.
\eea
This parameterization is convenient because all the divergences arises only 
from those terms containing $A_d$ or $D_d$.

%\be H_4(u)= 3\ln(1-u) + \frac{u}{2(1-u)} (6-3u-u^2)\ee

Thus the integrals
for the response function read (as usual $x=s/t$)
\bea
I_1(t,s)&=&\int_s^t \rmd t' R^0_{\q=0}(t,t')  I(t') R^0_{\q=0}(t',s) =
\int_s^t \rmd t'(t'/t)^{3/2} (s/t')^{3/2} 2 N_d r_d t'^{1-d/2}\nonumber\\
&=& 2 N_d r_d (s/t)^{3/2} \frac{t^{2-d/2} -s^{2-d/2}}{2-d/2}
= 2 N_d r_d (s/t)^{3/2}s^{\e/2}\left[-\ln x+\frac{1}{4}\e \ln^2 x+O(\e^2)\right],
\label{I1}
\eea
corresponding to Fig.~\ref{diagR}(1) and 
\bea
I_2(t,s)&=&\int_s^t \rmd t_1 \int_s^{t_1} \rmd s_1 R^0_{\q=0}(t,t_1) \sqrt{2} m_G(t_1)
B(t_1,s_1)\sqrt{2} m_G(s_1)  R^0_{\q=0}(s_1,s)\nonumber\\
%&=& \int_s^t \rmd t_1 \int_s^{t_1} \rmd s_1 (t_1/t)^{3/2} (3/t_1)^{1/2} (3/s_1)^{1/2}
%(s/s_1)^{3/2} 2(8\pi)^{-\dm} H_d(t_1,s_1)\nonumber\\
&=& 6(8\pi)^{-\dm}  \left(\frac{s}{t}\right)^{3/2} 
\int_s^t \rmd t_1\, t_1 \int_s^{t_1} \rmd s_1\, s_1^{-2} H_d(t_1,s_1)\,
\eea
corresponding to Fig.~\ref{diagR}(2).
Using the parameterization~\reff{para} for $H_d$,  
$I_2$ is reduced to five simple integrals whose sum is 
\bea
I_2(t,s)= \frac{6s^{\e/2}}{(8\pi)^{d/2}}  \left(\frac{s}{t}\right)^{3/2} 
&&\left[-\frac{2}{\e}\ln x -\frac{1-x}{2}+\frac{\pi^2}{3}+
3\ln(1-x)\left(\frac{1}{x}-1\right)+\frac{3}{2} \ln x
\right. \nonumber\\&&\left.
+\frac{1}{2}\ln^2 x -2\Li_2(x)+O(\e)\right]\,,
\label{I2}
\eea
with $x=s/t$ and $\Li_2(x) \equiv \sum_{n=1}^\infty x^n/n^2$.

\section{Feynman diagrams for the correlation function}
\label{appC}

There is only one new 1PI diagram for the correlation functions,
depicted in Fig.~\ref{diagC}(5) [we
define $t_< = \min\{t,t'\}$]:
\bea
B_{CC}(t,t')&=&\int (\rmd q) [C^0_\q(t,t')]^2=
\int (\rmd q) 
4(tt')^{-3}\rme^{-2\q^2(t+t')} \int_0^{t_<}\rmd t_1 t_1^3 \rme^{2\q^2t_1}
\int_0^{t_<}\rmd t_2 t_2^3 \rme^{2\q^2t_2}=
\nonumber\\&=&
\frac{4(tt')^{-3}}{(8\pi)^{d/2}}\int_0^{t_<} \rmd t_1\int_0^{t_<} \rmd t_2 
\frac{ t_1^3 t_2^3}{(t+t'-t_1-t_2)^{d/2}}\,.
\eea
This enter in the connected diagram represented in
Fig.~\ref{diagC}(5) [we assume $t>s$, given that $I_5(t,s) = I_5(s,t)$]:
\bea
I_5(t,s)&=&\int_0^s\rmd\t_1\int_0^t\rmd\t_2 \, R^0_{\q=0}(s,\t_1)\sqrt2 m_G(\t_1) 
B_{CC}(\t_1,\t_2)\sqrt{2} m_G(\t_2) R^0_{\q=0}(t,\t_2)\nonumber\\
&=& \int_0^s \rmd\t_1\int_0^s \rmd\t_2
(s/\t_1)^{-3/2}(3/\t_1)^{1/2}B_{CC}(\t_1,\t_2) (3/\t_2)^{1/2}(t/\t_2)^{-3/2}
\nonumber\\&&
+\int_0^s \rmd\t_1\int_s^t \rmd\t_2
(s/\t_1)^{-3/2}(3/\t_1)^{1/2}B_{CC}(\t_1,\t_2) (3/\t_2)^{1/2}(t/\t_2)^{-3/2}
\nonumber\\&=&
3(st)^{-3/2}\left[2\int_0^s \rmd\t_1\int_0^{\t_1} \rmd\t_2\, \t_1\t_2 B_{CC}(\t_1,\t_2) +
\int_0^s \rmd\t_1\int_s^t \rmd\t_2 \, \t_1\t_2 B_{CC}(\t_1,\t_2)\right]
\nonumber\\&=&
3(st)^{-3/2}[2 B+A]\,.
\eea
It is easy to realize that both $A$ and $B$ do not diverge for
$d\to4$. Therefore no dimensional regularization is required 
and one can calculate them directly in $d=4$:
\bea
A&=&4 (8\pi)^{-\dm}\int_0^s \rmd\t_1\int_s^t\rmd \t_2 
\frac{1}{(\t_1\t_2)^2}
\int_0^{\t_1}\rmd t_1\int_0^{\t_1}\rmd t_2
\frac{t_1^3t_2^3}{(\t_1+\t_2-t_1-t_2)^2} + O(\e)
\nonumber\\&=& 4(8\pi)^{-\dm}s^4 
\left[{\cal A}(s/t) -\frac{81}{200}-\frac{\pi^2}{40}+ \frac{24}{25}\ln2\right]
+ O(\e)\,,\\
B&=&4(8\pi)^{-\dm}\int_0^s \rmd\t_1\int_0^{\t_1}\rmd\t_2 \frac{1}{(\t_1\t_2)^2}
\int_0^{\t_2}\rmd t_1\int_0^{\t_2}\rmd t_2
\frac{t_1^3t_2^3}{(\t_1+\t_2-t_1-t_2)^2} + O(\e)
\nonumber\\&=&
4(8\pi)^{-\dm}s^4 \left(\frac{1447}{4800}+\frac{\pi^2}{160}
-\frac{12}{25}\ln2\right) + O(\e)\; ,
\eea
where
\be
{\cal A}(x) = \frac{1}{4} \int_0^x\rmd y_1 \, (y_1^4x^{-4}-1) \int_0^1\rmd x_1
\int_0^1\rmd x_2 \frac{x_1^3 x_2^3}{(1 + y_1^{-1} - x_1 - x_2)^2}\,, 
\label{Ax}
\ee
is a negative and monotonically decreasing function with ${\cal A}(x) =
- x^3/252 + O(x^4)$ and ${\cal A}(1) = \frac{1153}{3600} +
\frac{\pi^2}{30} - \frac{24}{25} \ln 2 \simeq -0.016$.
[Although ${\cal A}(x)$ can be expressed in
terms of elementary functions we do not report here its lengthy form.]
Thus
\be
I_5(t,s)= 3(st)^{-3/2} [A+2B]=
\frac{s(s/t)^{3/2}}{(8\pi)^{d/2}}\left[ 
\left(\frac{19}{8}-\frac{3\pi^2}{20}\right)+12{\cal A} (s/t)\right]+O(\e)\,.
\label{I5}
\ee

There are two diagrams involving $B_{RC}(t_1,s_1)$ [see
Fig.~\ref{diagC}(6)], namely
\bea
I_6(t,s)&=&
\int_0^t \rmd t_2 \int_{t_2}^t \rmd t_1\, R^0_{\q=0}(t,t_1) \sqrt{2} m_G(t_1) B_{RC}(t_1,t_2)
\sqrt{2} m_G(t_2) C^0_{\q=0}(t_2,s)+\nonumber\\&&
\int_0^s \rmd t_1 \int_{t_1}^s \rmd t_2\, C^0_{\q=0}(t,t_1) \sqrt{2} m_G(t_1) B_{RC}(t_2,t_1)
\sqrt{2} m_G(t_2) R^0_{\q=0}(s,t_2)\,,
\eea
that can be written as
\bea
I_6(t,s)&=
\frac{3}{2} (st)^{-3/2} &\left[
\int_0^s \rmd t_2 \, t_2^2\int_{t_2}^t \rmd \, t_1 t_1 B_{RC}(t_1,t_2)+
%\nonumber\\&+&
s^4 \int_s^t \rmd t_2\, t_2^{-2}\int_{t_2}^t \rmd t_1 \, t_1 B_{RC}(t_1,t_2)\right.
\nonumber\\&&\left.+
\int_0^s \rmd t_1 \, t_1^2\int_{t_1}^s \rmd t_2\, t_2 B_{RC}(t_2,t_1)\right]\,.
\eea
Using the parameterization of Eq.~(\ref{para}), each of the three integrals 
decomposes in five integrals that can be easily computed.
Therefore 
\be
I_6(t,s) = \frac{3}{2}(st)^{-3/2} [(A) + (B) + (C)] =
\frac{3(st)^{-3/2}}{(8\pi)^\dm} \sum_{a\in{A,B,C}} \sum_{i=1}^5 (a)_i
\ee 
where
{\small
\bea
(A)_1&=&\frac{A_d}{1+\e/2} 
\left[\frac{t^{1+\e/2}s^3}{3}- \frac{s^{4+\e/2}}{4+\e/2}\right]\\
(A)_2&=&-\frac{A_d}{1+\e/2}\left[
\frac{s^3 t}{3}-\frac{s^4}{4}+\frac{\e}{2} \left(
s\frac{9s^3-4s^2 t-6s t^2 -12t^3}{144}
\right.\right.\nonumber\\&&\left.\left.
+\frac{ t^4}{12} \ln t-
\frac{3s^4-4s^3t+t^4}{12}\ln (t-s)\right)
+O(\e^2)\right]\\
(A)_3&=&\frac{B_ds^4}{4}\left[\frac{1}{4}+\ln\frac{t}{s}\right]+O(\e)\\
(A)_4&=&C_d\left[\frac{s^4}{4}-\frac{s^5}{5t}\right]+O(\e)\\
(A)_5&=&D_d\left[\frac{s^4}{2\e}-\frac{3}{8}s^4 +\frac{s^5}{5t}-\frac{s^3t}{12}-
\frac{s^2t^2}{8}-\frac{st^3}{4}-\frac{s^4}{4}\ln\frac{t}{s}+
\frac{s^4-t^4}{4}\ln(t-s)+\frac{t^4}{4}\ln t \right]+O(\e)\\
(B)_1&=&\frac{s^4A_d}{1+\e/2}\left[t^{\e/2}\left(\frac{t}{s}-1\right)
-\frac{2}{\e}\left(t^{\e/2}-s^{\e/2}\right)\right] \\
(B)_2&=&-\frac{s^4A_d}{1+\e/2}\left[\frac{t}{s}-1
-\ln\frac{t}{s}+\frac{\e}{2}\left(\frac{\pi^2}{6}+\ln t \ln s-\ln^2t
-\ln\frac{t}{s}+\left(\frac{t}{s}-1\right)\ln (t-s)-\Li_2\frac{s}{t} \right)\right]\nonumber\\&&
+O(\e)\\
(B)_3&=&\frac{s^4B_d}{2}\ln^2 \frac{t}{s}\\
(B)_4&=&C_d s^4\left(\ln \frac{t}{s}-\frac{t-s}{t}\right) \\
(B)_5&=&D_d s^4\left[\frac{2}{\e}\ln\frac{t}{s}-\ln\frac{t}{s}+\frac{t-s}{t}-
\frac{1}{2}\ln^2\frac{t}{s} +\ln^2t-\frac{\pi^2}{6}-\ln t\ln s+\Li_2\frac{s}{t}
\right]+O(\e) \\
(C)_1&=&\frac{A_d}{1+\e/2}s^{4+\e/2}\left(\frac{1}{3}-\frac{1}{4+\e/2}\right)\\
(C)_2&=&-\frac{A_d}{1+\e/2}\frac{2  s^{4+\e/2}}{(2+\e/2)(3+\e/2)(4+\e/2)}\\
(C)_3&=&\frac{B_d}{16}s^4+O(\e)\\
(C)_4&=&\frac{C_d}{20}s^4+O(\e)\\
(C)_5&=&D_ds^4 \left[\frac{1}{2\e}-\frac{19}{30}+\frac{1}{4}\ln s\right]+O(\e)
\eea
}
Summing up one finds
\bea
I_6(t,s)&=&%\frac{3 (st)^{-3/2}}{(8\pi)^{d/2}}\sum_{i=1}^3\sum_{j=1}^5
	   %(ij) =
\frac{3 s^4 (st)^{-3/2}}{(8\pi)^{d/2}}\left[
\frac{1-2\ln x}{\e}+\frac{\pi^2}{3}-\frac{271}{60}\right.\nonumber\\&&+\left.
2\ln x + \ln t \left(\frac{1}{2}-\ln x\right)-\frac{\ln^2x}{2}+{\cal
B}(x)\right] +O(\e)\,.
\label{I6}
\eea
where 
\be
{\cal B}(x)=\frac{2 x}{5}-\frac{7}{2} \ln (1-x)-2
   \text{Li}_2(x)+\frac{31}{8}+\frac{4 \ln
   (1-x)}{x} 
- \frac{1}{2 x^4} \left[ \ln(1-x) + x + \frac{x^2}{2} + \frac{x^3}{3}\right]
%-\frac{1}{6 x}-\frac{1}{4 x^2}-\frac{1}{2
%   x^3}-\frac{\ln (1-x)}{2 x^4}\,
\label{Bx}
\ee
is a monotonically increasing function with ${\cal B}(x)= x^3/63 + O(x^4)$ and ${\cal B}(1) = \frac{403}{120}-\frac{\pi^2}{3}
\simeq 0.068$.
%Note that in the sum all the terms diverging for $x\to0$ cancel (except 
%the logarithm, obviously).

The diagram involving $I(t)$ is [see Fig.~\ref{diagC}(7)] 
\bea
I_7(t,s)&=&\int_0^s\rmd t'\, R^0_{\q=0}(t',s)I(t')C^0_{\q=0}(t,t')+
\int_0^t\rmd t'\, R^0_{\q=0}(t',t)I(t') C^0_{\q=0}(t',s)=\nonumber\\
&=& N_dr_d\left[2(st)^{-3/2}\int_0^s \rmd t'\, t'^{5-d/2}+
\frac{s^{5/2}}{t^{3/2}}\int_s^t \rmd t'\,t'^{1-d/2}\right]=\nonumber\\
&=&N_d r_d s \left(\frac{s}{t}\right)^{3/2}s^{\e/2}\left\{\frac{2}{4+\e/2}
+\frac{2}{\e}\left[\left(\frac{t}{s}\right)^{\e/2}-1\right]\right\}\,.
\label{I7}
\eea

Finally there is the contribution coming from the 
one-loop correction to the magnetization. 
At this order in perturbation theory it can be simply evaluated as
\be
C_{{\bf q}=0}^0(t,s)=2\int_0^s\rmd t_2\,
\exp\left[-\left(2\int_{t_2}^s \rmd t_1 \, m^2(t_1)
+\int_s^t \rmd t_1 \, m^2(t_1)\right)\right]\,.
\ee
Using the value of $m^2(t)$ given by Eq. (\ref{m1loop0}), we obtain
\bea
C_{{\bf q}=0}^0(t,s)&=&2\int_0^s \rmd t_2\, \exp\left\{-
3\int_{t_2}^s\frac{\rmd t_1}{t_1}\left(1-2\gt\frac{r_d}{1+\e/2}t_1^{\e/2}\right)-
\frac{3}{2}\int_s^t \frac{\rmd
t_1}{t_1}\left(1-2\gt\frac{r_d}{1+\e/2}t_1^{\e/2}\right)\right\}=\nonumber\\ 
&=&2\int_0^s \rmd t_2 \left(\frac{t_2}{s}\right)^3\left(\frac{s}{t}\right)^{3/2}
\left\{1+\frac{6\gt r_d}{1+\e/2}\left[\frac{2}{\e}(s^{\e/2}-t_2^{\e/2})+
\frac{1}{\e}\left(t^{\e/2}-s^{\e/2}\right)\right]+O(\gt^2)\right\}=
\nonumber\\&=&
\frac{1}{2}s\left(\frac{s}{t}\right)^{3/2}
\left\{1+\frac{6\gt r_d s^{\e/2}}{1+\e/2} \left[\frac{1}{4+\e/2}+
\frac{1}{\e}\left(\left(\frac{t}{s}\right)^{\e/2}-1\right)\right]\right\}\,.
\label{I8}
\eea

Summing up all the contributions according to Eq.~(\ref{Csum}) one finds
\be
C_{\q=0}(t,s)=\frac{1}{2}s\left(\frac{s}{t}\right)^{3/2}
\left[1+\gt \left(-\frac{719}{320}+\frac{37}{160}\pi^2-\frac{3}{8}\ln x
\right)+\gt f_C(s/t) +O(\gt^2,\e\gt)\right] \,,
\ee
with 
\bea
f_C(x)&=& \frac{3 x}{10}+12 {\cal A}(x)+ 3{\cal B}(x)-\frac{21}{8} \ln(1-x)-
\frac{3 \Li_2(x)}{2}+\frac{93}{32}+\nonumber\\ 
&&+\frac{3 \ln(1-x)}{x}
%%%
%-\frac{1}{8 x}-\frac{3}{16 x^2}-\frac{3}{8 x^3}-\frac{3 \ln (1-x)}{8 x^4}
-\frac{3}{8x^4}\left[ \ln(1-x) + x + \frac{x^2}{2} + \frac{x^3}{3}\right]
\label{fx}\,,
\eea
[$f_C(0) = 0$] where
${\cal A}(x)$ and ${\cal B}(x)$ are given by Eqs. (\ref{Ax}) 
and (\ref{Bx}).

\end{document}